\newif\ifmarkedits\markeditstrue
\documentclass[aps,prx,reprint,superscriptaddress,preprintnumbers,floatfix,nofootinbib,longbibliography]{revtex4-2}%
\makeatletter
\newcommand\showtitleinbib{{\escapechar=`\\ \immediate\write\@auxout{%
\csname citation{REVTEX42Control}\endcsname^^J%
\csname citation{apsrev42Control}\endcsname
}}}
\makeatother

\makeatletter
\let\origswitch@longtable\switch@longtable
\renewcommand\switch@longtable{\origswitch@longtable
\expandafter\let\csname longtable*\endcsname\@undefined
\expandafter\let\csname endlongtable*\endcsname\@undefined}
\makeatother

\usepackage{microtype}  
\usepackage{mathtools}  
\usepackage{amssymb,bm,bbm}
\usepackage{graphicx}
\usepackage{xcolor}
\usepackage[
  colorlinks=true,
  linkcolor=blue,
  citecolor=blue,
  filecolor=blue,
  urlcolor=blue,
]{hyperref}
\usepackage[capitalise]{cleveref}
\usepackage{orcidlink}

\usepackage{siunitx}
\usepackage{booktabs}  
\usepackage{slashed}
\usepackage{silence}
\usepackage{caption}
\usepackage{subcaption}
\usepackage{multirow}
\DeclareCaptionJustification{justified}{\leftskip=0pt\rightskip=0pt\parfillskip=0ptplus1fil\relax}
\usepackage{ytableau}
\ytableausetup{smalltableaux}
\newcommand\tydiagram[1]{\ytableausetup
  {aligntableaux=bottom}{\tiny\ydiagram{#1}}}
\usepackage{tikz}
\usetikzlibrary{tikzmark}
\usepackage{pgfplots}
\pgfplotsset{compat=1.18}

\usepackage{soul}

\newcommand{\<}{\langle}
\renewcommand{\>}{\rangle}
\renewcommand{\d}{\mathrm{d}}
\newcommand{\e}{\operatorname{e}}
\renewcommand{\i}{\mathrm{i}}
\newcommand{\tr}{\operatorname{tr}}

\newcommand{\hc}{\textnormal{h.c.}}

\newcommand{\Lg}{{L}}
\newcommand{\Rg}{{R}}

\newcommand{\Ul}{U}

\newcommand{\U}{\operatorname{U}}
\newcommand{\SU}{\operatorname{SU}}
\renewcommand{\O}{\operatorname{O}}
\newcommand{\SO}{\operatorname{SO}}

\newcommand{\su}{\mathfrak{su}}
\renewcommand{\so}{\mathfrak{so}}
\newcommand{\eff}{\mathrm{eff}}
\newcommand{\gauge}{\mathrm{gauge}}
\newcommand{\IR}{\mathrm{IR}}
\newcommand{\UV}{\mathrm{UV}}
\newcommand{\fr}{\mathrm{free}}
\newcommand{\wzw}{\mathrm{WZW}}
\renewcommand{\H}{\mathcal{H}}
\newcommand{\adj}{{\tiny
\begin{ytableau}
  \raisebox{2.5pt}[0pt][0pt]{\(\overline{\phantom{m}}\)} & 
\end{ytableau}}}
\newcommand{\anti}{\raisebox{0.5em}{\tiny\ydiagram{1,1}}}

\ifmarkedits\else\fi

\usepackage{relsize}

\usepackage[acronyms, nohypertypes={acronym}, nopostdot, style=super, nonumberlist, toc, shortcuts]{glossaries}
\glsenableentrycount
\setacronymstyle{long-sc-short}

\newacronym[longplural=quantum field theories]{QFT}{QFT}{quantum field theory}
\newacronym{RG}{RG}{Renormalization Group}
\newacronym{WZW}{WZW}{Wess-Zumino-Witten}
\newacronym{KS}{KS}{Kogut-Susskind}
\newacronym{NISQ}{NISQ}{noisy intermediate-scale quantum}
\newacronym{IR}{IR}{infrared}
\newacronym{QCD}{QCD}{quantum chromodynamics}

\begin{document}

\preprint{LA-UR-22-28886}
\title{Phases of 2d massless QCD with qubit regularization}
\author{Hanqing Liu\,\orcidlink{0000-0003-3544-6048}}
\email{hanqing.liu@lanl.gov}
\author{Tanmoy Bhattacharya\,\orcidlink{0000-0002-1060-652X}}
\email{tanmoy@lanl.gov}
\affiliation{Theoretical Division, Los Alamos National Laboratory, Los Alamos, New Mexico 87545, USA}
\author{Shailesh Chandrasekharan\,\orcidlink{0000-0002-3711-4998}}
\email{sch27@duke.edu}
\affiliation{Department of Physics, Box 90305, Duke University, Durham, North Carolina 27708, USA}
\author{Rajan Gupta\,\orcidlink{0000-0003-1784-3058}}
\email{rg@lanl.gov}
\affiliation{Theoretical Division, Los Alamos National Laboratory, Los Alamos, New Mexico 87545, USA}

\begin{abstract}
  We investigate the possibility of reproducing the continuum physics of 2d $\SU(N)$ gauge theory coupled to a single flavor of massless Dirac fermions using qubit regularization. The continuum theory is described by $N$ free fermions in the ultraviolet (UV) and a coset Wess-Zumino-Witten (WZW) model in the infrared (IR). In this work, we explore how well these features can be reproduced using the Kogut-Susskind Hamiltonian with a finite-dimensional link Hilbert space and a generalized Hubbard coupling. Using strong coupling expansions, we show that our model exhibits a gapped dimer phase and another phase described by a spin-chain. Furthermore, for $N=2$, using tensor network methods, we show that there is a second-order phase transition between these two phases. The critical theory at the transition can be understood as an $\SU(2)_1$ WZW model, using which we determine the phase diagram of our model quantitatively. Using the confinement properties of the model we argue how the UV physics of free fermions could also emerge, but may require further modifications to our model.
\end{abstract}

\maketitle

\section{Introduction}

The possibility of using quantum computers \cite{Ladd_2010} to understand quantum many-body physics and \acp{QFT} has opened up a new and exciting area of research at the crossroads of theoretical and computational physics \cite{Preskill:2018fag}.  Applications of quantum computing that span many areas of physics have been proposed \cite{Ayral:2023ron,Bassman:2021ljt,Bauer:2022hpo}. In particular, there have been extensive studies on formulating lattice gauge theories so that they can be studied efficiently on quantum computers 
\cite{Kaplan:2018vnj,Shaw:2020udc,Haase:2020kaj,Bauer:2021gek,Mathur:2016cko,Farrell:2022wyt,Farrell:2022vyh,Kan:2021xfc}. Quantum simulations of lattice gauge theories also have been proposed using both analog and digital quantum simulators, such as ultracold atoms in optical lattices \cite{Jaksch:1998zz,Banerjee:2012pg,Banerjee:2012xg,Tagliacozzo:2012vg,Tagliacozzo:2012df,Zohar:2011cw,Zohar:2012ay,Zohar:2012ts,Zohar:2012xf,Zohar:2013zla,Zohar:2015hwa,Ott:2020ycj,Gonzalez-Cuadra:2022hxt}, trapped ions \cite{Cirac:1995zz,Schindler:2013,Hauke:2013jga,Martinez:2016yna,Nguyen:2021hyk} and superconducting circuits \cite{You:2005,Marcos:2013aya,Belyansky:2023rgh}. There are several reviews available that discuss the progress that has been made \cite{Banuls:2019bmf,Bauer:2023qgm}.

Quantum computers work under a completely different computational paradigm than classical computers, requiring each degree of freedom of the theory to be encoded into qubits. Moreover, in the so-called \ac{NISQ} era \cite{Preskill:2018}, it is important to formulate the problems of interest so that they can be simulated efficiently using ``low-depth'' circuits. Simulating \acp{QFT} like QCD on a quantum computer to extract physically useful results will be particularly challenging. 

One major challenge of simulating a Hamiltonian lattice gauge theory with Lie groups on a quantum computer comes from its infinite-dimensional link Hilbert spaces. Traditionally, this theory is described by the \ac{KS} Hamiltonian \cite{Kogut:1974ag}, whose link Hilbert space corresponds to quantum particles moving on a group manifold like $\SU(N)$, and thus is infinite-dimensional. Hence one of the first steps of formulating a lattice gauge theory that can be simulated using a quantum computer requires a truncation of this infinite-dimensional Hilbert space to a finite-dimensional one. Recent research has focused on the myriad of ways to accomplish this truncation.

One popular approach is to restrict the particle's momentum, i.e., truncate in the group representation basis \cite{Byrnes:2005qx,Tagliacozzo:2014bta,Zohar:2014qma,Klco:2018kyo,Klco:2019evd,Ciavarella:2021nmj,Silvi:2019wnf,Rigobello:2023ype}. This approach manifestly preserves the gauge invariance, and the entire Hilbert space becomes block-diagonal in local symmetry sectors. The physical sector is the one that satisfies Gauss' law. Ideas of formulating the theory in this much smaller physical Hilbert space are also being actively explored \cite{Mathur:2004kr,Anishetty:2009ai,Mathur:2010wc,Anishetty:2014tta,Raychowdhury:2018osk,Raychowdhury:2019iki,Banuls:2017ena,Yao:2023pht,Muller:2023nnk}. An alternative approach is to discretize the particle's position, i.e., replace the Lie group by a discrete subgroup \cite{Zohar:2014qma,Bender:2018rdp,Lamm:2019bik,Alexandru:2019nsa,Hartung:2022hoz,Jakobs:2023lpp}. This approach maintains the diagonalizability of the Wilson plaquette operators, and is argued to be more suitable in the weak coupling regime \cite{Hartung:2022hoz,Jakobs:2023lpp}, where the continuum theory emerges. However, this is certainly not the only way that the continuum theory can emerge.

Other than truncating the Hilbert space, there are also other approaches that simply discard the infinite-dimensional Hilbert space, and replace it with a finite-dimensional one that is motivated by the symmetry of the theory. One of the most notable approaches in this direction is the D-theory introduced several years ago \cite{Brower:2003vy}. In the D-theory approach it was argued that by starting with a simple finite-dimensional local Hilbert space, but allowing it to grow through an extra dimension, we can regularize any \ac{QFT} including QCD \cite{Chandrasekharan:1996ih,Brower:1997ha,Wiese:2013uua}. Maintaining the symmetries allows one to study new types of gauge theories that have rich phase diagrams and interesting critical points \cite{Silvi:2016cas}. Development in this direction also includes deforming the symmetry algebra to a quantum group \cite{Zache:2023dko}, and the so-called fuzzy gauge theory \cite{Alexandru:2023qzd}.

The approach that we investigate in this work technically belongs to the first one, i.e., truncating the Hilbert space in the representation basis. However, we would like to emphasize that the philosophy behind our work focuses more on universal properties including symmetry. Many people working on this truncation approach regard the traditional Kogut-Susskind Hamiltonian as a target model, and assume many properties of this model still hold in the truncated model, even though these properties may be non-universal. For example, the continuum limit was explored in the limit $g^2 \rightarrow 0$ and $m \rightarrow 0$ in the lattice unit \cite{Banuls:2017ena,Rigobello:2023ype}. This perspective forces one to assume that the truncation of the Hilbert space introduces (non-universal) errors that need to be removed \cite{Ciavarella:2023mfc}.

In this work, we wish to present a different perspective. We view the formulation of \acp{QFT} using qubits as a new type of regularization of the \ac{QFT} which we refer to as qubit regularization \cite{Singh:2019uwd}. From this perspective, the \ac{KS} Hamiltonian is merely a starting point that has the same field content and symmetry as the continuum theory that we would like to reproduce. We do not impose any presumptions on this model, and allow the possibility of including other terms to reach the critical point of interest. In order to find a good qubit model, we first need to understand the symmetries, the \ac{RG} flow diagram and the phase structure of the continuum theory and isolate the relevant fixed points we wish to recover. We then explore qubit models that have the same symmetries and understand their phase diagrams. If there are critical points in the model, we study their properties and check whether they agree with the expectation from the continuum theory.

Our perspective is particularly helpful for studying QFTs using quantum computers, because given the limited resources during the NISQ era, it is important to find out what is the most efficient way to reach these critical points. There is evidence from spin models that we can even formulate continuum QFTs with asymptotic freedom using qubit models with a strictly finite-dimensional local Hilbert space \cite{Bhattacharya:2020gpm,Maiti:2023kpn}, which would be surprising if one believed that asymptotic freedom requires an infinite-dimensional local Hilbert space. 

This work focuses on reproducing some of the universal features of the $\SU(N)$ gauge theory in 2d with one flavor of massless Dirac fermions. This is partly motivated by the possibility of efficient classical simulation using the density matrix renormalization group (DMRG) algorithm \cite{White:1992zz} in the Hamiltonian formalism that is shared by quantum simulation, although tensor network methods in higher dimensions are also being actively studied \cite{Liu:2013nsa,xie2012coarse,Meurice:2020pxc,Akiyama:2022eip,Akiyama:2023hvt,Emonts:2018puo}. At the same time, QCD in 2d also provides a fertile ground for exploring properties of QCD in 4d, including confinement and chiral symmetry breaking. Here, we explore an extended qubit model motivated by the flow diagrams of the continuum theory. For general $N$, we use strong coupling expansions to study the different phases of the theory and their properties. For $N = 2$, we use DMRG to construct a more quantitative phase diagram and to investigate the critical properties of the model. Our findings offer a new perspective on the phase structure of 2d massless QCD and demonstrate the utility of qubit regularization in this context. 

This work is organized as follows. In \cref{sec:continuum}, we review the continuum theory of 2d QCD, where we analyze the phases of the theory using non-abelian bosonization and RG flow. We identify the RG flow between two fixed points, a free-fermion fixed point and a WZW fixed point, and compute the beta function near them. In \cref{sec:lattice-traditional}, we review the phase diagram of the traditional lattice model. In \cref{sec:qubit}, we review the structure of the link Hilbert space and introduce our qubit regularized model. In \cref{sec:strong}, we use strong coupling expansion of our qubit model to determine the phases and confinement properties. We argue that many of the results we derive are also valid in the traditional model. Finally in \cref{sec:critical}, we use the tensor network method to determine the phase boundary, study the critical physics, and compare our qubit model with the continuum theory and the traditional lattice model. We show that our qubit model can recover the WZW fixed point in a clean way, but the physics near the free fermion requires further investigation.

\section{The continuum theory and bosonization}\label{sec:continuum}
In this section, we review the continuum physics of 2d QCD, i.e., $\SU(N)$ Yang-Mills theory coupled to a single flavor of massless Dirac fermions in the fundamental representation. The Euclidean Lagrangian is given by 
\begin{align}\label{eq:free-Lagrangian}
  \mathcal{L}_0 = \frac{1}{2\tilde g^2} \tr F^2 + \bar\psi^\alpha \i\slashed D \psi^\alpha,
\end{align}
where $\alpha=1,\cdots,N$ is the color index. The first term is the kinetic term of the gauge field, and the second term is the kinetic term of the Dirac fermions as well as their coupling to the gauge field through the covariant derivative $\slashed D$. Here we use $\tilde g^2$ for the gauge coupling in the continuum theory to distinguish it from the bare coupling $g^2$ that we will introduce later on the lattice. 

In the following, we will review the physics of \cref{eq:free-Lagrangian} using non-Abelian bosonization. We will then extend the model with four-fermion current interactions, which naturally arise when we regularize the theory on a lattice. We then calculate its $\beta$ functions and discuss the RG flow.


\subsection{Bosonization}

The physics of $\mathcal{L}_0$ in 2d is best understood through bosonization \cite{Witten:1983ar,Knizhnik:1984nr,Delmastro:2021otj}. When $\tilde{g}^2=0$, the fermions are free and the fermion kinetic term exhibits $\U(N)_L \times \U(N)_R$ chiral symmetry. The low-energy physics of these massless fermions can be described by the $\U(N)_1$ WZW model \cite{Witten:1983ar}. When $\tilde g^2 > 0$, the $\SU(N)$ vector symmetry is gauged and the $\U(N)_L \times \U(N)_R$ chiral symmetry is broken to $\U(1)_L \times \U(1)_R$, and the low-energy physics is described by the $\U(N)_1\big/\SU(N)_1 \cong \U(1)_N$ coset WZW model \cite{Delmastro:2021otj}. There is another way to bosonize the theory when each Dirac fermion is viewed as two Majorana fermions,
\begin{align}
  \psi^\alpha = \frac{1}{\sqrt{2}}(\xi^{2\alpha-1} - \i \xi^{2\alpha}).
\end{align}
From this perspective, the free theory actually has an $\O(2N)_L \times \O(2N)_R$ symmetry and can be bosonized to the $\SO(2N)_1$ WZW model. When the $\SU(N)$ subgroup is gauged, the model is described by the $\SO(2N)_1\big/\SU(N)_1$ coset WZW model. This perspective is particularly useful for $N=2$, because in this case $\SO(4) \cong \SU(2)_c \times \SU(2)_s$, where the subscripts $c$ and $s$ mean ``charge'' and ``spin'' respectively. After gauging the $\SU(2)_s$ symmetry, the resulting coset WZW model is simply the $\SU(2)_1$ WZW model in the charge sector. For a detailed discussion of the spin and charge sectors, see \cite{Liu:2020ygc}.

In general, a $G_k \big/ H_{k'}$ WZW model has central charge $c = c(G_k) - c(H_{k'})$. In our case, at the free-fermion fixed point, i.e., $\tilde{g}^2=0$, we have $c_\fr = c(\U(N)_1) = c(\SO(2N)_1) = N$. When $\tilde g^2 > 0$, the low energy theory is described by the coset WZW model, and using $c(\SU(N)_1) = N-1$, we consistently obtain the same central charge $c_\wzw = 1$ from both $\U(N)_1\big/\SU(N)_1$ and $\SO(2N)_1\big/\SU(N)_1$ coset WZW models. Since a model is gapped if and only if its central charge $c = 0$ \cite{Delmastro:2021otj}, we know the coset WZW models here are gapless. According to the $c$ theorem \cite{Zamolodchikov:1986gt}, we expect there to be a flow from the free-fermion fixed point ($c_{\rm free} = N$) to the WZW fixed point ($c_{\rm WZW} = 1$), which will be discussed in \cref{subsec:beta}. 

\subsection{Interactions}
\label{subsec:interactions}
When $\tilde g^2 > 0$, the Lagrangian $\mathcal{L}_0$ has a $\U(1)_L \times \U(1)_R$ chiral symmetry and an $\SU(N)$ gauge symmetry for $N\geq 3$. However, when the model is regularized on a lattice using staggered fermions, the original $\U(1)_L \times \U(1)_R$ chiral symmetry is typically broken to the diagonal $\U(1)$ vector symmetry and a $\mathbb{Z}_2$ chiral symmetry, which is realized as a translation-by-one-site symmetry. While the discrete chiral symmetry prevents any fermion mass terms from being generated, it does allow marginal current-current couplings of the form $\lambda^a J_L^a J_R^a$, where
\begin{align}\label{eq:J}
  J_{L,R}^a := \frac{1}{2}\xi_{L,R}^T T^a \xi_{L,R},
\end{align}
and $T^a: a= 1, \cdots, N(2N-1)$ is a basis of the Lie algebra $\so(2N)$ and is normalized as $\tr(T^a T^b) = \delta^{ab}$ (see \cref{app:Lie} for our convention of the Lie algebras). Therefore, in order to describe the continuum theory emerging from the lattice, we must consider the following extended Lagrangian,
\begin{align}\label{eq:Lagrangian}
  \mathcal{L} = \frac{1}{2\tilde g^2} \tr F^2 + \bar\psi^\alpha \i\slashed D \psi^\alpha + \sum_a \lambda^a J_L^a J_R^a.
\end{align}
Using the $\U(1)$ vector global symmetry and the $\SU(N)$ gauge symmetry, we have further constraints on the $N(2N-1)$ couplings $\lambda^a$. Under the $\SU(N)$ gauge transformations, the $N(2N-1)$ currents form three real irreducible representations (irreps): a one-dimensional trivial representation $\circ$, an $N(N-1)$-dimensional anti-symmetric representation\footnote{The anti-symmetric representation further decomposes into two irreps over $\mathbb{C}$, but it is irreducible over $\mathbb{R}$.} $\anti$, and an $(N^2-1)$-dimensional adjoint representation $\adj$. The $\U(1)$ vector symmetry does not mix different representations. Thus, for $N\geq 3$ there are three independent couplings, which will be denoted as $\lambda_0$, $\lambda_{\tilde{c}}$, and $\lambda_s$ respectively. The coupling $\lambda_0$ is the well-known Thirring coupling.

The case of $N=2$ is special. Due to the isomorphism $\SO(4) \cong \SU(2)_c \times \SU(2)_s$, \cref{eq:J} has to preserve the $\SU(2)_c$ global symmetry and the $\SU(2)_s$ gauge symmetry, leading to only two independent couplings labeled by $\lambda_c$ and $\lambda_s$ respectively. This case can be viewed as a special case of $N\geq 3$, where $\lambda_0 = \lambda_{\tilde{c}} = \lambda_c$.

Here we note that none of the current interactions will be generated in the IR if we begin with a theory that has all the symmetries of $\mathcal{L}_0$. Clearly $\lambda_0$ and $\lambda_{\tilde{c}}$ (or $\lambda_c$ in the $N=2$ case) break the $\U(1)_L \times \U(1)_R$ chiral symmetry. It may appear that $\lambda_s$ can be generated, because after gauging the $\SU(N)$ vector symmetry, the chiral $\SU(N)$ symmetry becomes anomalous and does not prevent the $\lambda_s$ term from being generated. However, bosonization results indicate that in the IR limit, the $\su(N)_1$ current algebra is quotiented out, i.e., the currents associated with the $\SU(N)$ symmetry are set to zero, effectively removing the \(\lambda_s\) terms.

\subsection{Beta functions and flow diagrams}\label{subsec:beta}

Let us now understand the RG flow of the couplings in ${\cal L}$. We first focus on the flow diagram of the couplings $\lambda^a$ at the free-fermion fixed point, i.e., $\tilde{g}^2 = 0$. In 2d, these couplings are all marginal near the free-fermion fixed point, and the leading terms in the beta function of the general interaction $\lambda^a J_L^a J_R^a$ are related to the OPE coefficients \cite{gogolin2004bosonization,Affleck:1990iv,Affleck:1988zj}, and was calculated explicitly in \cite{Liu:2022},
\begin{align}\label{eq:beta-generic}
  \frac{\d\lambda^a}{\d\ln\mu} = -\frac{1}{4\pi} \sum_{b,c} (f^{abc})^2 \lambda^b \lambda^c,
\end{align}
where $f^{abc}$ is the structure constant\footnote{Note that there is an additional factor of $\frac{1}{4}$ compared to \cite{Liu:2022} due to the factor $\frac{1}{2}$ in the definition of $J_{L,R}^a$ in \cref{eq:J}. The normalization of $T^a$ only affects the beta function through the structure constant.}  of $\so(2N)$.
In the special case of $\lambda^a = \lambda$ for all $a$, we have
\begin{align}\label{eq:beta-isotropic}
  \frac{\d\lambda}{\d\ln\mu} = -\frac{1}{4\pi} \lambda^2 \sum_{b,c} (f^{abc})^2 = -\frac{N-1}{2\pi} \lambda^2 ,
\end{align}
where we have used the fact that $\sum_{b,c} (f^{abc})^2$ is related to the Killing form and is equal to $2N-2$ for $\so(2N)$ (see \cref{app:Lie}). In this case, the flow always stays on the line $\lambda^a = \lambda$ because it is protected by the $\O(2N)$ vector symmetry.

On the other hand, if we only require the $\U(1) \times \SU(N)$ symmetry, then as discussed in the previous section, there are three independent couplings $\lambda_0$, $\lambda_{\tilde{c}}$ and $\lambda_s$. From \cref{eq:beta-generic}, we know that these couplings satisfy
\begin{align}
  \frac{\d\lambda_A}{\d\ln\mu} = -\frac{1}{4\pi} \sum_{\{B,C\}} F^{ABC} \lambda_B \lambda_C,
\end{align}
where $A, B, C \in \{0, \tilde{c}, s\}$, and
\begin{align}
  F^{ABC} = F^{ACB} = (2-\delta_{BC})\sum_{b\in B,c \in C} (f^{abc})^2.
\end{align}
$F^{ABC}$ is independent of the choice of $a \in A$ due to the $\SU(N)$ symmetry. It turns out that only five independent $F^{ABC}$ are non-zero\footnote{This can be easily checked using the following facts about the Lie algebra $\so(2N)$: (a) $\mathfrak{l} := \circ \oplus \adj$ and $\mathfrak{p} := \anti$ form a Cartan decomposition of $\so(2N)$; (b) $[\circ, \adj]=0$; and (c) $\so(2N)$ and $\su(N)$ are semisimple. }: $F^{0\tilde{c}\tilde{c}}$, $F^{\tilde{c}0\tilde{c}}$, $F^{\tilde{c}s\tilde{c}}$, $F^{sss}$, $F^{s\tilde{c}\tilde{c}}$. Furthermore, \cref{eq:beta-isotropic} implies that $F^{0\tilde{c}\tilde{c}} = 2N-2$, $F^{\tilde{c}0\tilde{c}} + F^{\tilde{c}s\tilde{c}} = 2N-2$, $F^{sss} + F^{s\tilde{c}\tilde{c}} = 2N-2$. It can be calculated that $F^{\tilde{c}0\tilde{c}} = \frac{4}{N}$, while $F^{sss} = N$ is related to the Killing form of $\su(N)$. Putting these results together, we find that $\lambda_0$, $\lambda_{\tilde{c}}$ and $\lambda_s$ satisfy the following beta functions
\begin{subequations}\label{eq:beta-g=0-N>3}
  \begin{align}
    \frac{\d\lambda_0}{\d\ln\mu} &= -\frac{N-1}{2\pi} \lambda_{\tilde{c}}^2, \\
    \frac{\d\lambda_{\tilde{c}}}{\d\ln\mu} &= -\frac{1}{2\pi} \big(\frac{2}{N} \lambda_0 \lambda_{\tilde{c}} + (N-1-\frac{2}{N}) \lambda_s \lambda_{\tilde{c}} \big), \\
    \frac{\d\lambda_s}{\d\ln\mu} &= -\frac{1}{4\pi} \big(N \lambda_s^2 + (N-2) \lambda_{\tilde{c}}^2\big).
  \end{align}
\end{subequations}
When $N=2$, the $\SU(2)_c$ symmetry requires that $\lambda_0 = \lambda_{\tilde{c}} = \lambda_c$. Then \cref{eq:beta-g=0-N>3} is simplified to 
\begin{align}\label{eq:beta-g=0-N=2}
  \frac{\d\lambda_{c,s}}{\d\ln\mu} &= -\frac{1}{2\pi} \lambda_{c,s}^2.
\end{align}
Notice that $\lambda_c$ and $\lambda_s$ are irrelevant when they are less than zero, and the free-fermion fixed point becomes an IR fixed point in this.

Now let us consider the effects of $\tilde g^2 > 0$ on the flow diagram. In order to get a full picture of the flow diagram, we will need to compute the above $\beta$ functions including $\tilde{g}^2$. Instead of doing this, we use some heuristic arguments to motivate a possible flow diagram. First, we note that in 2d, the gauge coupling $\tilde{g}$ is a relevant direction. Near the free-fermion fixed point, it has dimension of mass. Further, as we discussed in the previous subsection, as $\tilde{g}^2$ flows to the IR, the $\lambda_s$ sector effectively decouples from the theory. Another way to see this is that $\lambda_s$ and $\tilde{g}^2$ belong to the same symmetry sector, and $\lambda_s$ is marginal while $\tilde{g}^2$ is highly relevant. This suggests that the flow of $\lambda_s$ becomes a part of the $\tilde{g}^2$ flow, and the $\lambda_s$ term disappears in the IR, except at $\tilde{g}^2=0$. With these observations, let us now discuss the flow diagram including $\tilde{g}^2$.



\begin{figure}[htb]
\centering
\includegraphics[width=0.4\textwidth]{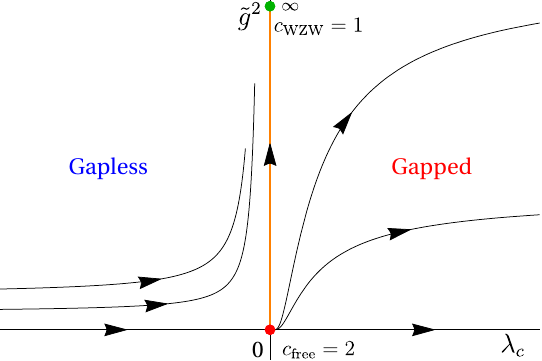}
\caption{Flow diagram in the $\lambda_c$ and $\tilde g^2$ plane for $N=2$. 
  The origin $\tilde{g}^2=\lambda_c=0$ is the red free-fermion fixed point with $c_\mathrm{free}=2$, while $\tilde{g}^2=\infty$ and $\lambda_c=0$ is the green WZW fixed point with $c_\mathrm{WZW}=1$. When $\tilde{g}^2>0$, the flow diagram shows that the Lagrangian ${\cal L}$ describes two phases depending on the sign of $\lambda_c$: a gapped phase for $\lambda_c > 0$, and a gapless phase that flows to the WZW fixed point for $\lambda_c < 0$. The Lagrangian ${\cal L}_0$ lives on the orange line defined by $\lambda_c = 0$, which separates the two phases. When $\tilde{g}^2=0$, the plot assumes $\lambda_s = \lambda_c$, and when $\lambda_s = \lambda_c<0$, it flows to the free-fermion fixed point.}
  \label{fig:flow-diagram}
\end{figure}

In \cref{fig:flow-diagram}, we plot the flow diagram in the $\lambda_c$ and $\tilde g^2$ plane, where for simplicity and clarity, we focus on the case of $N=2$. We see that there are two fixed points: a red free-fermion fixed point at $\tilde{g}^2=\lambda_c=0$ with central charge $c_\mathrm{free}=N=2$, and a green WZW fixed point at $\tilde{g}^2=\infty$ and $\lambda_c=0$ with central charge $c_\mathrm{WZW}=1$. Since there is a flow from the free-fermion fixed point to the WZW fixed point, we will also refer to them as the UV fixed point and the IR fixed point, respectively. When $\tilde g^2>0$ and $\lambda_c>0$, both couplings flow to infinity, and we have a gapped phase. On the other hand when $\tilde g^2>0$ and $\lambda_c<0$, the system flows to the green WZW fixed point, and we have a gapless phase. The orange line indicates the phase boundary between the gapped and the gapless phase and corresponds to the continuum theory with Lagrangian ${\cal L}_0$. When $\tilde g^2=0$ we assume $\lambda_s=\lambda_c$ so that the system flows to the free-fermion fixed point when $\lambda_c \leq 0$, and is gapped when $\lambda_c > 0$.


\begin{figure}[htb]
  \centering
  \includegraphics[width=0.3\textwidth]{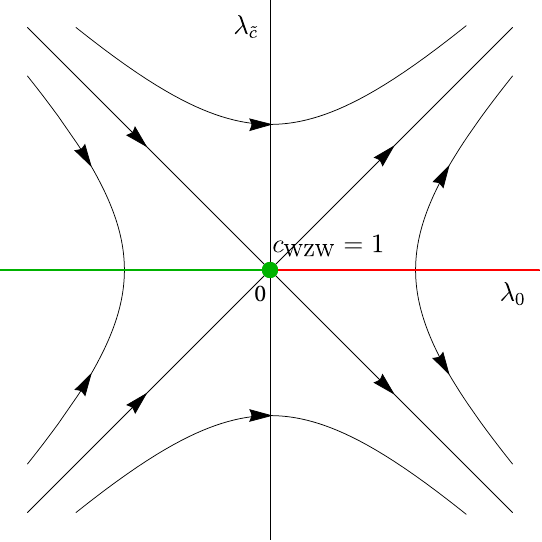}
  \caption{Flow diagram in the $\lambda_0$ and $\lambda_{\tilde{c}}$ plane near the WZW fixed point when $N\geq 3$. The flow lines are hyperbolas, and the lines $\lambda_0 = \pm \sqrt{N(N-1)/2} \lambda_{\tilde{c}}$ define the separatrix. At $\lambda_{\tilde{c}} = 0$ there is a line of fixed points which are attractive when $\lambda_{\tilde{c}} < 0$, colored with green, and repulsive when $\lambda_{\tilde{c}} \geq 0$, colored with red. The origin corresponds to the free QCD theory, which flows to the green WZW fixed point in the IR. When $N=2$, the flow is restricted to the line $\lambda_0 = \lambda_{\tilde c}$.}
  \label{fig:BKT-flow-diagram}
\end{figure}

When $N\geq 3$, there is also an interesting flow diagram near the WZW fixed point, which appears as the IR fixed point when $\tilde g^2 > 0$. In this IR limit, the currents associated with the $\SU(N)$ symmetry vanish, which effectively removes the terms containing \(F^{\tilde{c}s\tilde{c}}\), \(F^{sss}\), and \(F^{s\tilde{c}\tilde{c}}\) from the flow equations. Then the beta functions in \cref{eq:beta-g=0-N>3} are simplified to
\begin{subequations}\label{eq:beta-N>=3}
  \begin{align}
    \frac{\d\lambda_0}{\d\ln\mu} &= -\frac{N-1}{2\pi} \lambda_{\tilde{c}}^2, \\
    \frac{\d\lambda_{\tilde{c}}}{\d\ln\mu} &= -\frac{1}{N\pi} \lambda_0 \lambda_{\tilde{c}},
  \end{align}
\end{subequations}
which leads to the Berezinski-Kosterlitz-Thouless (BKT) flow diagram plotted in \cref{fig:BKT-flow-diagram}.
The flow lines are hyperbolas because
\begin{align}
  \frac{\d}{\d\ln\mu} \big(\frac{2}{N}\lambda_0^2 - (N-1)\lambda_{\tilde{c}}^2\big) &= 0,
\end{align}
and the lines $\lambda_0 = \pm \sqrt{N(N-1)/2}\, \lambda_{\tilde{c}}$ define the separatrix. There is a line of fixed points at $\lambda_{\tilde{c}}=0$. This is expected because $\lambda_0$ is the Thirring coupling, which is known to be conformal. When $\lambda_0<0$, these fixed points are attractive, and are the IR fixed points of the gapless phase; when $\lambda_0>0$, they are repulsive, and separate two gapped phases. When $N=2$, we have $\lambda_0 = \lambda_{\tilde{c}} = \lambda_c$, which sits right on one of the separatrices.


In both \cref{fig:flow-diagram,fig:BKT-flow-diagram}, the gapped phases can be further constrained by a mixed anomaly between the $\U(1)$ fermion number symmetry and the $\mathbb{Z}_2$ chiral symmetry. Using 't Hooft anomaly matching \cite{tHooft:1979rat}, we expect IR phases to be either gapless or gapped with spontaneous $\mathbb{Z}_2$ chiral symmetry breaking \cite{Lieb:1961fr,Jian:2017skd,McGreevy:2022oyu}---which, on the lattice, corresponds to spontaneous breaking of the translation-by-one-site symmetry. In \cref{subsec:phase,sec:critical}, we find that using strong coupling expansion, the ground states in the gapped phases indeed break the lattice translation symmetry and are degenerate.

\section{The traditional lattice model}\label{sec:lattice-traditional}

In this section, we will review the traditional lattice model that is used to reproduce the continuum physics of ${\cal L}_0$ defined in \cref{eq:free-Lagrangian}. In particular, we will introduce the Hamiltonian and explain its symmetries. We will then argue how we can also extend the model to study the physics of ${\cal L}$ defined in \cref{eq:Lagrangian}. We will show how the continuum limit of the lattice theory emerges and discuss the free-fermion fixed point

\subsection{The Kogut-Susskind Hamiltonian}
\label{subsec:KS-Hamiltonian}
One popular approach to regularize the continuum theory ${\cal L}_0$ in \cref{eq:free-Lagrangian} on the lattice is the Kogut-Susskind Hamiltonian \cite{Kogut:1974ag},
\begin{align}\label{eq:KS}
  H_0 &= \frac{g^2}{2} \sum_{\<i,j\>} \big( L_{ij}^{a2} + R_{ij}^{a2} \big) + t\sum_{\<i, j\>} (c_i^{\alpha\dagger} U_{ij}^{\alpha\beta} c_j^\beta + \hc ),\!\! 
\end{align}
where $i = 1, \cdots, L$ labels the lattice sites, and $ij$ refers to the link connecting the sites $i$ and $j$. The first term is the electric field term, where $L_{ij}^a, R_{ij}^a: a = 1, \cdots, N^2-1$ are the electric field operators that generate left and right $\SU(N)$ gauge transformations on the link $ij$; the second term is the gauged fermion hopping term. $c_i^{\alpha\dagger}, c_i^\alpha: \alpha = 1, \cdots, N$ are fermion creation and annihilation operators on site $i$, and $U_{ij}^{\alpha\beta}: \alpha,\beta = 1, \cdots, N$ are the Wilson link operators that lives in the fundamental representation of $\SU(N)$, and implement parallel transport on the link $ij$.

This Hamiltonian has a $\U(1)$ global symmetry generated by the fermion number operator
\begin{align}
  \hat N := \sum_{i=1}^L n_i =  \sum_{i=1}^L \sum_{\alpha=1}^N c_i^{\alpha\dagger} c_i^\alpha,
\end{align}
and an $\SU(N)$ gauge symmetry generated by the gauge transformation generators
\begin{align}
  G_i^a := c_i^{\alpha\dagger} T_{\alpha\beta}^a c_i^\beta + L_{i,i+1}^a + R_{i-1,i}^a
\end{align}
on each site $i$. The operators obey the following commutation relations that lead to \([G_i^a, H] = 0\),
\begin{subequations}\label{eq:symalg-lgt}
  \begin{align}
    [L_i^a, L_j^b] &= \i f^{abc} L_i^c \delta_{ij}, ~~
    [R_i^a, R_j^b] = \i f^{abc} R_i^c \delta_{ij}, \\
    [L_i^a, R_j^b] &= 0, \\
    [L_i^a, U_{jk}^{\alpha\beta}] &= -T^a_{\alpha\gamma} U_{jk}^{\gamma\beta} \delta_{ij}, 
    [R_i^a, U_{jk}^{\alpha\beta}] = U_{jk}^{\alpha\gamma} T^a_{\gamma\beta} \delta_{ik}, \\
    [G_i^a, c_j^\alpha] &= -T_{\alpha\beta}^a c_j^\beta \delta_{ij}, ~~
    [G_i^a, c_j^{\alpha\dagger}] = c_j^{\beta\dagger} T_{\beta\alpha}^a \delta_{ij},
  \end{align}
\end{subequations}
where $T^a, a=1, \cdots,N^2-1$ are the generators of $\SU(N)$ in the fundamental representation. In addition to the symmetry algebra in \cref{eq:symalg-lgt}, in the traditional approach we also have the commutation relations
\begin{align}\label{eq:U-commute}
  [U_{ij}^{\alpha\beta}, U_{kl}^{\gamma\delta}] = 0,
\end{align}
which are not needed to show $[G_i^a, H] = 0$. As we will see in the next section, it is the freedom in this commutator that we will utilize to achieve qubit regularization while preserving the symmetries.

Due to the gauge symmetry, the Hamiltonian of a lattice gauge theory block diagonalizes into sectors that can be labeled by the gauge charges (irreps of $\SU(N)$) on each site. The physical sector is spanned by all physical states $|\psi\>$ that are invariant under gauge transformations, i.e., $G_i^a |\psi\> = 0$ for all $i$ and $a$, which is known as the Gauss's law constraint. This constraint is equivalent to $\sum_a (G_i^a)^2 |\psi\> = 0$ for all $i$, i.e., $|\psi\>$ carries a trivial representation of $\SU(N)$ on every lattice site.

When $g^2=0$ we can attach a Wilson line to $c_i^\alpha$, and define a new set of fermion annihilation operators $d_i^\alpha$ as
\begin{align}
  d_i^\alpha := U_{12}^{\alpha\alpha_2}U_{23}^{\alpha_2\alpha_3} \cdots U_{i-1,i}^{\alpha_{i-1}\alpha_i} c_i^{\alpha_i}.
\end{align}
Then using the unitarity of the $U_{ij}^{\alpha\beta}$ operators, i.e., $\sum_\alpha (U_{ij}^{\alpha\beta})^\dagger U_{ij}^{\alpha\gamma}$, we obtain
\begin{align}
  H_0 &= t\sum_{\<i, j\>} (d_i^{\alpha\dagger} d_j^\beta + \hc ),
\end{align}
which is the free-fermion Hamiltonian. Note that while we can still perform this redefinition at $g^2 >0$, the resulting theory does not yield a free-fermion theory because the fermion hopping term does not commute with the electric field term, due to the $U_{ij}^{\alpha\beta}$ operators contained in the definition of $d_i^\alpha$.

Thus, we see that the free-fermion fixed point of the continuum theory is reproduced by the Hamiltonian in \cref{eq:KS} at $g^2=0$. For this reason, the continuum limit of \cref{eq:KS} is obtained when we tune the bare lattice coupling $g^2 \rightarrow 0$, where the full flow from UV to IR of the Lagrangian ${\cal L}_0$ is recovered.

\subsection{The extended Hamiltonian}
\label{subsec:full-Hamiltonian}

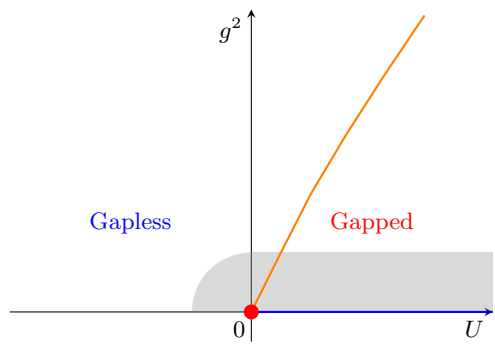
\begin{figure}[htb]
  \centering
  \begin{tikzpicture}
  \begin{axis}[
    axis lines=middle,
    xlabel={$U$},
    ylabel={$g^2$},
    xlabel style={at={(ticklabel* cs:1)}, anchor=north east},
    ylabel style={at={(ticklabel* cs:1)}, anchor=north east},
    xmin=-2, xmax=2,
    ymin=-0.5, ymax=5.1,
    xtick=\empty, ytick=\empty,
    width=8cm,
    height=6cm
    ]

    \fill[gray, opacity=0.3, rounded corners=0mm] (axis cs:0,0) rectangle (axis cs:2,1);

    \begin{scope}
      \clip (axis cs:0,0) rectangle (axis cs:-2,2); 
      \fill[gray, opacity=0.3] (axis cs:0,0) circle (0.79cm); 
    \end{scope}
    
    \addplot[orange, thick] coordinates {
      (0, 0)
      (0.24484963, 1)
      (0.49582378, 2)
      (0.78887172, 3)
      (1.1046641, 4)
      (1.43500144, 5)
    };

    \draw[blue, thick] (axis cs:0,0) -- (axis cs:2,0);

    \node[red, circle, fill, inner sep=2pt] at (axis cs:0,0) {};

    \node at (axis cs:-0.1, -0.3) {0};
    \node[red] at (axis cs:1, 1.5) {Gapped};
    \node[blue] at (axis cs:-1, 1.5) {Gapless};
  \end{axis}
\end{tikzpicture}
  \caption{Schematic phase diagram of the traditional lattice model for $N=2$. The orange line is the boundary between a gapped phase and a gapless phase. The red dot indicates the free-fermion critical point. The blue line at $g^2=0$ and $U > 0$ describes a line of critical points whose long-distance physics is described by the $\SU(2)_s$ WZW model. As we will argue in \cref{subsec:phase} using strong coupling expansion, the unshaded region can be accessed easily by simple qubit regularization schemes, while accessing the gray-shaded requires more careful numerical study.}
  \label{fig:phase-diagram-traditional}
\end{figure}

When $g^2\neq 0$, the lattice Hamiltonian in \cref{eq:KS} is not guaranteed to be on the $\lambda_c=0$ line in \cref{fig:flow-diagram}. As we discussed in \cref{subsec:interactions}, a marginal current-current coupling is allowed by the symmetry of the lattice model, and in general it will be generated. One can ask whether this current-current coupling can be captured by some lattice interactions, by tuning which we can get back to the $\lambda_c=0$ line. Such lattice interactions must not break the $\U(1) \times \SU(N)$ symmetry of $H_0$ (or $\SU(2)_c \times \SU(2)_s$ symmetry in the $N=2$ case). One such term is the generalized Hubbard term\footnote{Note that the coupling $U$ we introduce here should not be confused with gauge link operator $U_{ij}^{\alpha\beta}$, which always appears with indices.}
\begin{align}
  H_U = -U \sum_in_i(N-n_i),   
\end{align}
where $n_i$ is the total fermion number operator on site $i$, which clearly preserves the $\U(1) \times \SU(N)$ symmetry. When $N=2$ this term is reduced to the usual Hubbard coupling at half-filling
\begin{align}
  n_i(2-n_i) = 2(n_i^1 - \frac{1}{2}) (n_i^2 - \frac{1}{2}) - \frac{1}{2},
\end{align}
which is known to preserve the $\SU(2)_c \times \SU(2)_s$ symmetry of $H_0$. Another choice is 
\begin{align}
  H_V = V \sum_{\<i,j\>} (n_i-\frac{N}{2})(n_j-\frac{N}{2}) ,  
\end{align}
which is a generalization of the coupling in the $t-V$ model \cite{Huffman:2013mla,Huffman:2016nrx}, and can be used to reproduce the Thirring coupling as argued in \cite{Liu:2020ygc}. When $N\geq 3$, we know the continuum theory $\mathcal{L}$ has two independent interactions, and we may need both $H_U$ and $H_V$ to explore the full phase diagram. On the other hand, when $N=2$, there is only one independent interaction and we only need $H_U$, because $H_V$ does not preserve the $\SU(2)_c$ symmetry. Since our numerical work is focused on $N=2$, in the following we will only study the extended Hamiltonian
\begin{align}\label{eq:ks-with-Hubbard}
  H &= H_0 + H_U.
\end{align}
The phase diagram of this extended model in the $U$ and  $g^2$ plane is schematically plotted in \cref{fig:phase-diagram-traditional} for $N=2$. When $N\geq 3$, the same phase diagram might be reproduced, but we may need a combination of both $H_U$ and $H_V$ so that the combined coupling moves along a suitable direction through the origin in the RG flow diagram in \cref{fig:BKT-flow-diagram}.

The phases in the region $g^2/t \gg 1$ or $-U/t \gg 1$ is justified by the strong coupling analysis explained in \cref{subsec:strong-traditional}. The phase transition indicated by the orange line is second order at large $g^2$ based on the numerical work we discuss in \cref{sec:critical}. Notice that the orange phase boundary between the gapped and the gapless phase must meet the red point at $g^2 = U = 0$, because when $g^2 = 0$ the fermions are free of the gauge interactions, and we know that Hubbard coupling $U$ is perturbatively proportional to the current-current coupling $\lambda$ \cite{Affleck:1988zj,Liu:2020ygc}. Therefore we know at $g^2 = U = 0$ must be a critical point, which corresponds to the free-fermion fixed point of the continuum theory. In addition, at $g^2=0$ and $U>0$, the model is fermions with no gauge interactions, but with a Hubbard interaction, which is also gapless and can be described by the $\SU(2)_s$ WZW model. This is represented by the thick blue line on the positive $U$ axis. Comparing the phase diagram in \cref{fig:phase-diagram-traditional} and the flow diagram in \cref{fig:flow-diagram}, the physics of the continuum theory $\mathcal{L}_0$ can be reproduced by first approaching the orange line for every $g^2 > 0$ and then following it towards $g^2=0$.

As we will discuss in the remainder of this paper, the infinite-dimensional link Hilbert space on which the Hamiltonian \cref{eq:ks-with-Hubbard} acts can be drastically reduced while still preserving most of the phase diagram of \cref{fig:phase-diagram-traditional}. This is the so-called qubit regularization approach. 

\section{Qubit regularization}\label{sec:qubit}

When we introduced the Hamiltonians \cref{eq:KS} and \cref{eq:ks-with-Hubbard} in the previous section, we did not mention the Hilbert space on which they act. In this section, we review the structure of the gauge-link Hilbert space, and explain the idea of qubit regularization as introduced in \cite{Liu:2021tef}.

\subsection{Traditional link Hilbert space}

Since we focus on a single link in this section, we will suppress the spatial location of the link in the discussion below. In traditional lattice gauge theories, the link Hilbert space is the space of square-integrable functions on the group manifold $\SU(N)$ which we will denote as $\H$. 
One basis of $\H$ is the ``position'' basis, which consists of states $|g\>: g\in \SU(N)$. One can view $|g\>$ as the position of the quantum particle on the group space. These states satisfy the orthonormal and complete relations 
\begin{align}
  \< g|g'\> = \delta(g-g'),\quad \int \d g |g\>\< g| = 1,
\end{align}
where $\d g$ is the Haar measure on the group manifold. The link operators $U^{\alpha\beta}$ can also be viewed as ``position'' operators that satisfy 
\begin{align}
\label{eq:uong}
  U^{\alpha\beta}|g\> = D^f_{\alpha\beta}(g)|g\>,
\end{align}
where $D^f(g)$ is an $N \times N$ matrix corresponding to the fundamental representation of $\SU(N)$. Thus in the position basis all link operators $U^{\alpha\beta}$ are diagonal. 

In contrast to $U^{\alpha\beta}$, the operators $\Lg^a$ and $\Rg^a$ are not (block-)diagonal in the position basis, since they are the ``momentum'' operators that generate the group of left-translations $\SU(N)_L$ and right-translations $\SU(N)_R$ on the group manifold $\SU(N)$. Under these left and right translations $(h_L, h_R) \in \SU(N)_L \times \SU(N)_R$, we have $|g\> \mapsto |h_L^{-1} g h_R\>$. There is a ``momentum'' basis of $\H$, which consists of states 
\begin{align}
\label{eq:mombasis}
|D^\lambda_{\alpha\beta}\>: \lambda \in \widehat{\SU(N)}, \alpha,\beta \in \dim\lambda,    
\end{align}
where $D^\lambda_{\alpha\beta}$ is a function on the group manifold $\SU(N)$, corresponding to the matrix element in the irrep $\lambda$, and $\widehat{\SU(N)}$ denotes the set of inequivalent irreps of the $\SU(N)$ group. This basis is related to the ``position'' basis through
\begin{align}
  \<g|D^\lambda_{\alpha\beta}\> = \sqrt{d_\lambda} D^\lambda_{\alpha\beta}(g),
\end{align}
where $D^\lambda_{\alpha\beta}(g)$ is the matrix element of $g$ in irrep $\lambda$, and $d_\lambda$ is the dimension of $\lambda$. This relation is known as a generalized Fourier transform, and hence the analogy between position and momentum. Peter-Weyl theorem \cite{Harlow:2018tng} states that the ``momentum'' basis is also orthonormal and complete, i.e.,
\begin{align}
  \<D^\lambda_{\alpha\beta}|D^{\lambda'}_{\alpha'\beta'}\> = \delta_{\lambda\lambda'} \delta_{\alpha\alpha'} \delta_{\beta\beta'},~ \sum_{\lambda} \sum_{\alpha,\beta} |D^{\lambda}_{\alpha\beta}\> \<D^{\lambda}_{\alpha\beta}| = 1,
\end{align}
and thus $\H$ can be decomposed as,
\begin{align}\label{eq:hs-lgt}
  \H \cong \bigoplus_{\lambda\in\widehat{\SU(N)}} V_\lambda \otimes V_\lambda^*,
\end{align}
where $V_\lambda \otimes V_\lambda^* = \mathrm{span}\{|D^\lambda_{\alpha\beta}\>: \alpha,\beta \in \dim\lambda\}$.

The operators $L^a$ and $R^a$ are block diagonal in this basis with matrix elements
\begin{align}\label{eq:LRmat-suN}
  \<D^\lambda_{\alpha\beta}|\Lg^a|D^\lambda_{\gamma\delta}\> &= -(T^a_\lambda)_{\gamma\alpha}\delta_{\beta\delta},\nonumber\\
  \<D^\lambda_{\alpha\beta}|\Rg^a|D^\lambda_{\gamma\delta}\> &= (T^a_\lambda)_{\beta\delta}\delta_{\alpha\gamma},
\end{align}
where $T^a_\lambda$ are the generators of $\SU(N)$ in the irrep $\lambda$. On the other hand, the operators $\Ul_{\alpha\beta}$ connect states in different irreps,
\begin{align}\label{eq:U-action}
\Ul_{\alpha\beta} |D^\lambda_{\gamma\delta}\> &= \int \d g \sqrt{d_\lambda} D^\lambda_{\gamma\delta}(g) D^f_{\alpha\beta}(g) |g\>,
\end{align}
where $D^\lambda_{\gamma\delta}(g)D^f_{\alpha\beta}(g)$ is an element in the tensor product representation $\lambda \otimes f$. Since all irreps of $\SU(N)$ can be labeled by the Young diagrams and the fundamental representation is labeled by a single box $\tydiagram{1}$, the representation $\lambda \otimes f$ decomposes into a sum of irreps that corresponds to adding a single box to the Young diagram of $\lambda$ consistent with the branching rules. 

\subsection{Regularized link Hilbert space}
\label{subsec:qubit}

Since the traditional link Hilbert space $\H$ is infinite-dimensional, a traditional lattice gauge theory cannot be formulated on a quantum computer directly without some form of regularization. This leads to the idea of qubit regularization, where we construct lattice gauge theories with a finite-dimensional link Hilbert space that is in the same universality class as the traditional theory.

In order to remain in the same universality class, we preserve the symmetries of the Hamiltonian, including the gauge symmetry, i.e. $[H, G_i^a] = 0$ for all lattice sites $i$ and $a = 1, \cdots, N^2-1$. A simple way to achieve this is to project the full Hilbert space $\H$ to a subspace $\H_Q$,
\begin{align}\label{eq:HQ}
  \H_Q := \bigoplus_{\lambda \in Q} V_\lambda \otimes V_\lambda^*,
\end{align}
where $Q$ is a finite subset of $\widehat{\SU(N)}$. We note that after this truncation, the relation $[U_{ij}^{\alpha\beta}, U_{kl}^{\gamma\delta}] = 0$ in \cref{eq:U-commute} and the unitarity of $U$ no longer holds.

As pointed out in \cite{Liu:2021tef}, one particularly simple choice with nice properties is all the irreps labeled by Young diagrams with one column and at most $N-1$ rows\footnote{These representations, except for the trivial representation, are also known as the ``fundamental representations'' of the Lie algebra $\su(N)$ to mathematicians, because the highest weights of these representations correspond to the fundamental weights of $\su(N)$. We avoid this terminology since in the physics literature, the fundamental representation usually refers to \(\ydiagram{1}\).}, i.e.,
\begin{align}
  Q = \{\circ,~ \ydiagram{1},~ \ydiagram{1,1},~ \ydiagram{1,1,1},~ \cdots,~ \overline{\ydiagram{1,1}},~ \overline{\ydiagram{1}}\},
  \label{eq:qr}
\end{align}
where the overline on $k$ boxes indicates a dual representation with $N-k$ boxes. This regularization scheme was also studied recently by other groups in the case of $\SU(3)$, and was referred to as hardcore lattice QCD \cite{Silvi:2019wnf,Rigobello:2023ype}. 

There are several reasons for this choice. First, due to the gluon screening, it is believed that string tensions in confining theories at large distance are dictated by $N$-alities, i.e., the number of boxes modulo $N$, rather than the detailed representations, and our $Q$ contains all $N$-alities. Second, each representation in $Q$ has the smallest quadratic Casimir among representations with the same $N$-ality, i.e., they minimize the electric field energy $\frac{g^2}{2} \big( L^{a2} + R^{a2} \big)$, which is also responsible for the string tensions. Third, as will be argued in \cref{sec:strong}, in the $g^2/t \rightarrow \infty$ limit, this qubit regularization scheme has the same physics as traditional theory. This suggests that any quantum critical points in the qubit regularized models that arise at weaker couplings can have analogs in the traditional theory. Finally, these representations are precisely the representations that fermions transform under. Therefore it is easy to form gauge invariant operators by combining fermions and gauge links.

Another qubit regularization scheme, that can be useful if the test quarks are in the fundamental representation and its dual, and we are interested in deep IR physics, can be chosen as
\begin{align}
  \bar Q = \{\circ,~ \ydiagram{1},~ \overline{\ydiagram{1}}\} =  \{\mathbf{1},~ \mathbf{N},~ \mathbf{\bar N}\}.
\end{align}
This regularization scheme $\bar Q$ coincides with $Q$ for $\SU(2)$ and $\SU(3)$, but for larger values of $N$ it is much simpler and could be of interest if one can design Hamiltonians with desired quantum critical points.

Qubit regularization modifies the definition of the link operators $U^{\alpha\beta}$ to $U^{\alpha\beta}_Q$ and the left and right generators $\Lg^a$ and $\Rg^a$ to $\Lg^a_Q$ and $\Rg^a_Q$ by projecting these operators to the link Hilbert space $\H_Q$ defined in \cref{eq:HQ}. Thus, we simply add a subscript $Q$ to represent the projection, and our qubit Hamiltonian can be written as
\begin{align}\label{eq:qubit-model}
  H_Q &= H_{0,Q} + H_U.
\end{align}
which is formally the same as \cref{eq:ks-with-Hubbard}. In the following, this model will be simply referred to as the ``qubit model''.

Here we remark an important distinction of the qubit model \cref{eq:qubit-model} as compared to the traditional model \cref{eq:ks-with-Hubbard} regarding the range that the parameter $g^2$ can take. In the traditional model, $g^2$ is strictly non-negative, because otherwise its spectrum is unbounded from below. However, in the qubit model, $g^2$ is allowed to be negative, because the link Hilbert space is finite-dimensional, making its spectrum always bounded.

\section{Strong coupling expansion}
\label{sec:strong}

In this section, we explore the possible phase diagram and confinement properties of our qubit model defined in \cref{eq:qubit-model} for arbitrary $N$ using strong coupling expansions, where the hopping term $t$ is small compared to $U$ or $g^2$.

In these limits, at leading order we can set $t=0$ and the Hamiltonian is diagonal in the representation basis of the link Hilbert space and the fermion occupation number basis on the sites. The representation of the link Hilbert space can be labeled by the number of boxes $0\leq k < N$ in the Young diagram. Therefore each basis state of the full Hilbert space can be labeled as $|\{k_i\}, \{n_i\}, \alpha\>$, where $k_i$ is the number of boxes on the link connecting site $i$ and $i+1$, $n_i$ is the number of fermions on site $i$, and $\alpha$ is all other quantum numbers that distinguish degenerate states. The energy eigenvalue equation then reads
\begin{align}
  H_Q|\{k_i\}, \{n_i\}, \alpha\> = E_{\{k_i\}, \{n_i\}}|\{k_i\}, \{n_i\}, \alpha\>.
\end{align}

$E_{\{k_i\}, \{n_i\}}$ can be determined by noticing that $L_{ij}^{a2}$ and $R_{ij}^{a2}$ are proportional to quadratic Casimir operators and have eigenvalues $N c_2(k_i)$, which is briefly reviewed in \cref{app:Lie}. Therefore the energy eigenvalues of the basis states are given by
\begin{align}
  E_{\{k_i\}, \{n_i\}} = g^2\sum_i \frac{N+1}{2N}k_i(N-k_i) - U \sum_in_i(N-n_i).
\label{eq:scen}
\end{align}
When the full Hilbert space is restricted to the physical Hilbert space by imposing Gauss's law, the physical states have to satisfy
\begin{align}
  n_i + k_{i-1} - k_i \equiv 0 \mod N \quad \text{ for } i = 1, \cdots L.
\end{align}
Given $k_0$ and $\{n_i\}$, this constraint can be solved as
\begin{align}
  k_i = \Big(k_0 + \sum_{j=1}^i n_j \mod N\Big).
  \label{eq:fixedlinks}
\end{align}
On the other hand, we can also eliminate all the $n_i$'s by
\begin{align}
  n_i \equiv k_i - k_{i-1} \mod N \quad \text{ for } i = 1, \cdots, L.
\end{align}
Since $H_U$ is invariant under $n_i \leftrightarrow N-n_i$, it can be rewritten in terms of $k_i$ as
\begin{align}
H_U = -U \sum_i |k_i - k_{i-1}|(N-|k_i - k_{i-1}|).
\end{align}

Armed with these results, we can analyze the ground state structures and confinement properties in various strong coupling limits.

\subsection{Phase diagram}
\label{subsec:phase}
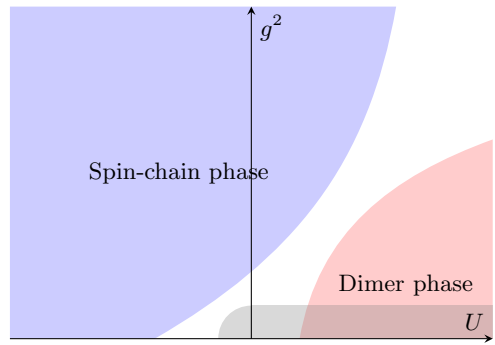
\begin{figure}[htb]
  \centering
  \scalebox{1}{\begin{tikzpicture}
  \begin{axis}[
    axis lines=middle,
    axis on top,
    xlabel={$U$},
    ylabel={$g^2$},
    xmin=-5, xmax=5,
    ymin=0, ymax=5,
    xtick=\empty, ytick=\empty,
    width=8cm,
    height=6cm  
    ]
    \fill[blue!20] (axis cs: -2, 0) -- (axis cs: -5, 0) -- (axis cs: -5, 5) -- (axis cs: 3, 5) to[out=-100, in=30] cycle;
    \node[align=center] at (axis cs: -1.5, 2.5) {Spin-chain phase};
    \fill[red!20] (axis cs: 5, 0) -- (axis cs: 1, 0) to[out=80, in=-160] (axis cs: 5, 3) -- cycle;
    \node[align=center] at (axis cs: 3.2, 0.8) {Dimer phase};
    
    \fill[gray, opacity=0.3, rounded corners=0mm] (axis cs:0,0) rectangle (axis cs:5,0.5);
    \begin{scope}
      \clip (axis cs:0,0) rectangle (axis cs:-2,2); 
      \fill[gray, opacity=0.3] (axis cs:0,0) circle (0.44cm); 
    \end{scope}
  \end{axis}
\end{tikzpicture}
}
  \caption{Schematic phase diagram of the qubit model in the strong coupling limits. The blue region indicates a spin-chain phase, which is gapless for $N=2$, but gapped for $N\geq 3$; the red region indicates a dimer phase, which is generally gapped; the gray shaded region is where the phases might be different between the qubit model and the traditional model.}
  \label{fig:phase-diagram-strong}
\end{figure}

In this subsection, we will show that in the strong coupling limits, our model has a spin-chain phase and a dimer phase. In particular, as is shown in \cref{fig:phase-diagram-strong}, when $-U/t \gg 1$ or $g^2/t \gg 1$, the qubit model is in a spin-chain phase: when $N=2$, the spin-chain phase is described by a Heisenberg spin chain, which is gapless; when $N\geq 3$, it is described by an XXZ spin chain in the gapped phase. When $U/t \gg 1$, the qubit model is in a dimer phase, which is gapped in general. Furthermore, when the system is gapped, we show that the ground states always break the translation-by-one-site symmetry, and are always degenerate, which is expected from the 't Hooft anomaly matching argument that we discussed at the end of \cref{subsec:beta}. As we will see, most of this strong coupling analysis also applies to the traditional model, except for the gray shaded region in \cref{fig:phase-diagram-strong}.

\subsubsection{Spin-chain phase}\label{subsec:spin-chain}

When $-U/t \gg 1$, the sites prefer to be empty or filled with $N$ fermions, i.e. $n_i =0$ or $N$ for all $i$. Then Gauss's law enforces that $k_i = k_0$ for all $i$ for some fixed $0 \leq k_0 < N$. When $g^2/t \gg 1$, links prefer to be in the trivial representation, i.e. $k_i = 0$. Then too Gauss's law requires that all sites be either empty or filled for all $i$. Thus the limits $-U/t \gg 1$ and $g^2/t \gg 1$ are compatible with each other, and imposing both limits is possible, which is highlighted as the blue region in \cref{fig:phase-diagram-strong}. Let us now discuss how the qubit model simplifies in these limits.

First notice that the gauge field and fermions satisfy Gauss's law on their own and are decoupled. The physics in the gauge field is simple: it is a constant. In the fermion sector, only singlet states are allowed on each site, i.e., $n_i = 0$ or $N$. Thus, the effective Hilbert space dimension is $2^L$, the same as a spin-half chain. A typical state can be pictorially represented as 
\begin{align}
  \begin{tikzpicture}[baseline={([yshift=-.8ex]current bounding box.center)}]
  \fill[blue!10] (1,0.25) ellipse (0.3 and 0.4);
  \fill[blue!10] (4,0.25) ellipse (0.3 and 0.4);
  \fill[blue!10] (6,0.25) ellipse (0.3 and 0.4);
  {\color{blue!80!black}
    \draw(0.1,0) -- (0.9,0);
    \draw(1.1,0) -- (1.9,0);
    \draw(2.1,0) -- (2.9,0);
    \draw(3.1,0) -- (3.9,0);
    \draw(4.1,0) -- (4.9,0);
    \draw(5.1,0) -- (5.9,0);
    \draw(6.1,0) -- (6.9,0);
  }
  {\color{red!80!black}
    \node at (0.9,-0.05) [above] {$\uparrow$} ;
    \node at (1.1,-0.05) [above] {$\downarrow$} ;
    \node at (3.9,-0.05) [above] {$\uparrow$} ;
    \node at (4.1,-0.05) [above] {$\downarrow$} ;
    \node at (5.9,-0.05) [above] {$\uparrow$} ;
    \node at (6.1,-0.05) [above] {$\downarrow$} ;
    \node[fill, circle, inner sep=1pt] at (1, 0) {};
    \node[fill, circle, inner sep=1pt] at (2, 0) {};
    \node[fill, circle, inner sep=1pt] at (3, 0) {};
    \node[fill, circle, inner sep=1pt] at (4, 0) {};
    \node[fill, circle, inner sep=1pt] at (5, 0) {};
    \node[fill, circle, inner sep=1pt] at (6, 0) {};
  }
\end{tikzpicture}
,
\end{align}
for $N=2$. The red dots represent the lattice sites, the blue bubbles represent fermion singlets with $n_i=2$, and the blue lines between the sites represent $k_i=0$. 

All the $2^L$ states of the Hilbert space are degenerate at the zeroth order of the fermion hopping term $t$. However, this degeneracy is lifted by $t$, and in \cref{app:spin-chain}, we show that by performing a perturbation calculation in $t$, we obtain the following effective XXZ spin chain Hamiltonian,
\begin{align}
  H_\eff = \sum_{\<i,j\>} J_\perp (X_i X_j + Y_i Y_j) + J_z (Z_i Z_j - 1) ,
\end{align}
where $X_i, Y_i, Z_i$ are Pauli operators on site $i$, and
\begin{align}
  J_\perp &= (-1)^{N-1} \frac{N}{2(N-1)!} \frac{t^N }{(\frac{N+1}{2N} g^2 - 2U)^{N-1}}, \nonumber\\
  J_z &= \frac{N}{2(N-1)} \frac{t^2}{\frac{N+1}{2N} g^2 - 2U}.
\end{align}

The mapping of our qubit model to the spin chain helps us understand its phase structures. It is well-known that the phase of an XXZ spin chain depends on the sign of $J_z$ and the ratio $\Delta := J_z/J_\perp$ \cite{giamarchi2004quantum}. In our case, $J_z$ is always positive, and the model is in some anti-ferromagnetic phase. When $N=2$, we have $\Delta = -1$, and $H_\eff$ is a Heisenberg anti-ferromagnetic spin chain, which is gapless with central charge $1$, and is consistent with the gapless phase of the continuum model in \cref{fig:flow-diagram}. When $N \geq 3$, $J_\perp$ is of higher order than $J_z$, which means in the strong coupling limit, $|\Delta|\gg 1$ and $H_\eff$ flows to a gapped N\'eel phase, where the translation-by-one-site symmetry is spontaneously broken as expected from 't Hooft anomaly matching argument. We also note that the symmetry of the spin chain is consistent with the global symmetry of our Hamiltonian $H_Q$, which is $\SU(2)_c$ for $N=2$ and $\U(1)$ for $N \geq 3$. 

\begin{figure}[htb]
  \centering
  \includegraphics[width=0.3\textwidth]{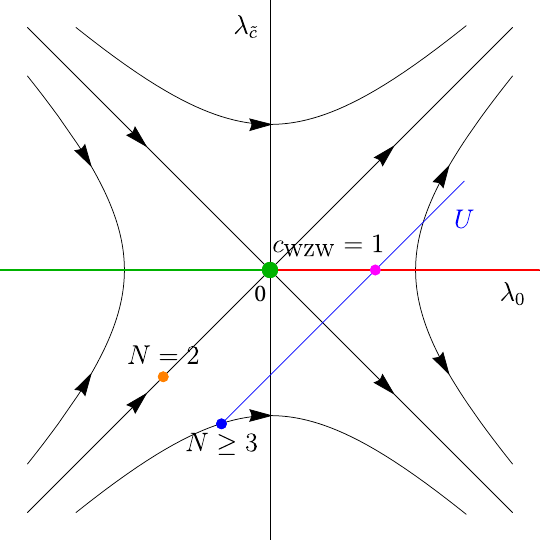}
  \caption{The location of the qubit model in the strong coupling limit in the flow diagram discussed in \cref{fig:BKT-flow-diagram}. The orange dot indicates the location of the qubit model in the limits $-U/t \gg 1$ or $g^2/t \gg 1$ for $N=2$. Increasing $U$ will move it along the $\lambda_0 = \lambda_{\tilde c}$ line closer to the WZW fixed point, and eventually undergo a phase transition at the fixed point into the dimerized massive phase. The blue dot indicates the location of the qubit model in the limits $-U/t \gg 1$ or $g^2/t \gg 1$ for $N\geq 3$, which is in a gapped phase. Increasing $U$ will move it along the blue line. When it intersects with the red line at the pink dot, one would observe a phase transition to another gapped phase characterized by the dimerized ground states.}
  \label{fig:BKT-flow-diagram-Hubbard}
\end{figure}

In \cref{fig:BKT-flow-diagram-Hubbard}, we qualitatively show the location of our qubit model in the flow diagram of $\lambda_0$ and $\lambda_{\tilde c}$ in \cref{fig:BKT-flow-diagram}, and discuss the role of the Hubbard coupling $U$. When $N=2$, the qubit model must sit on the line $\lambda_0 = \lambda_{\tilde c}$ due to the $\SU(2)_c$ symmetry, and in the limits $-U/t \gg 1$ or $g^2/t \gg 1$, it must be in the third quadrant in order to flow to the WZW fixed point. When $N \geq 3$, in the strong coupling limit we know the qubit model is gapped, therefore it sits around the blue dot and flows to the fixed point in the fourth quadrant at infinity. As we will show in the next subsection, when $U/t \gg 1$, the qubit model is also gapped but with dimerized ground states, and we expect the qubit model to be controlled by a different fixed point, possibly the one in the first quadrant. Therefore the generalized Hubbard coupling $U$ could induce a phase transition between two massive phases.

\subsubsection{Dimer phase}\label{subsec:dimer}
When $U/t \gg 1$, highlighted as the red region in \cref{fig:phase-diagram-strong}, the physics is very different from what we discussed above, since in this case the fermions repel each other and cannot form gauge singlets by themselves. We also have to treat even and odd $N$ separately.


We will first consider the case of even $N$. In this case, the sites all prefer half-filling $n_i = N/2$. Then due to Gauss's law, the links have to be in the alternating pattern of $k_0$ and $(k_0 + \frac{N}{2})\mod N$. If $g^2=0$, the link energy is always zero independent of the value of $k_0$ and the ground state is $N$-fold degenerate, corresponding to the $N$ distinct values that $k_0$ can take. On the other hand, if $g^2>0$, the ground state is $2$-fold degenerate, because the energy of two nearest links
\begin{align}
  k_0(N-k_0) + \big(\frac{N}{2} + k_0\big) \big(\frac{N}{2} - k_0\big)
\end{align}
is minimized at $k_0 = 0$ and $k_0 = \frac{N}{2}$. For both $g^2=0$ and $g^2>0$, all other states are separated by a gap of order $U$. There are two ground states each of which is dimerized breaking the translation-by-one-site symmetry and the system is gapped.

In the $U/t\rightarrow\infty$ limit, the two ground states can be written down explicitly as a tensor product of dimers
\begin{align}\label{eq:SUN-dimer}
  \bigotimes_{i=1}^{L/2} \frac{1}{\binom{N}{N/2}} \Big(|\alpha_1 & \cdots \alpha_{\frac{N}{2}}\>_{2i-1} |D^{[\frac{N}{2}]}_{[\alpha_1 \cdots \alpha_{\frac{N}{2}}][\gamma_1 \cdots \gamma_{\frac{N}{2}}]}\>_{2i-1, 2i} \nonumber\\
  &\varepsilon^{\gamma_1 \cdots \gamma_{\frac{N}{2}}\beta_1 \cdots \beta_{\frac{N}{2}}}|\beta_1 \cdots \beta_{\frac{N}{2}}\>_{2i}\Big)
\end{align}
where
\begin{align}
  |\alpha_1 \cdots \alpha_{\frac{N}{2}}\>_i := c_i^{\alpha_{\frac{N}{2}}\dagger} \cdots c_i^{\alpha_1\dagger} |0\>_i,
\end{align}
$\alpha$ are assumed to satisfy $\alpha_1 < \cdots < \alpha_{\frac{N}{2}}$, the same is true for $\beta$, $\gamma$ etc., and $\varepsilon^{\alpha_1\alpha_2 \cdots }$ is the Levi-Civita symbol. With periodic boundary conditions, translating \cref{eq:SUN-dimer} by one site leads to the other ground state. For $N=2$, \cref{eq:SUN-dimer} is simplified to
\begin{align}\label{eq:SU2-dimer}
  \bigotimes_{i=1}^{L/2} \frac{1}{2} |\alpha\>_{2i-1} |D^{}_{\alpha\gamma}\>_{2i-1, 2i} \varepsilon^{\gamma\beta}|\beta\>_{2i},
\end{align}
which can be pictorially represented as
\begin{align}
  \begin{tikzpicture}[baseline={([yshift=-.8ex]current bounding box.center)}]
  \fill[blue!10] (1.5,0.25) ellipse (0.7 and 0.4);
  \fill[blue!10] (3.5,0.25) ellipse (0.7 and 0.4);
  \fill[blue!10] (5.5,0.25) ellipse (0.7 and 0.4);
  {\color{blue!80!black}
    \draw(0,0) -- (1,0);
    \draw(1,0.5) -- (2,0.5);
    \draw(2,0) -- (3,0);
    \draw(3,0.5) -- (4,0.5);
    \draw(4,0) -- (5,0);
    \draw(5,0.5) -- (6,0.5);
    \draw(6,0) -- (7,0);
  }
  {\color{red!80!black}
    \node at (1,-0.05) [above] {$\uparrow$} ;
    \node at (2,-0.05) [above] {$\downarrow$} ;
    \node at (3,-0.05) [above] {$\uparrow$} ;
    \node at (4,-0.05) [above] {$\downarrow$} ;
    \node at (5,-0.05) [above] {$\uparrow$} ;
    \node at (6,-0.05) [above] {$\downarrow$} ;
  }
\end{tikzpicture}
,
\end{align}
where the red arrows represent fermions, a lower blue line represents a gauge link in the state $k=0$, an upper blue line represents a gauge link in the state $k=1$, and a bubble represents one dimer. Notice the direction of the arrows for the fermions is just for display, because the fermion colors should be contracted to the gauge link indices to make a gauge singlet on each site.

\begin{figure}[htb]
  \centering \includegraphics[width=0.48\textwidth]{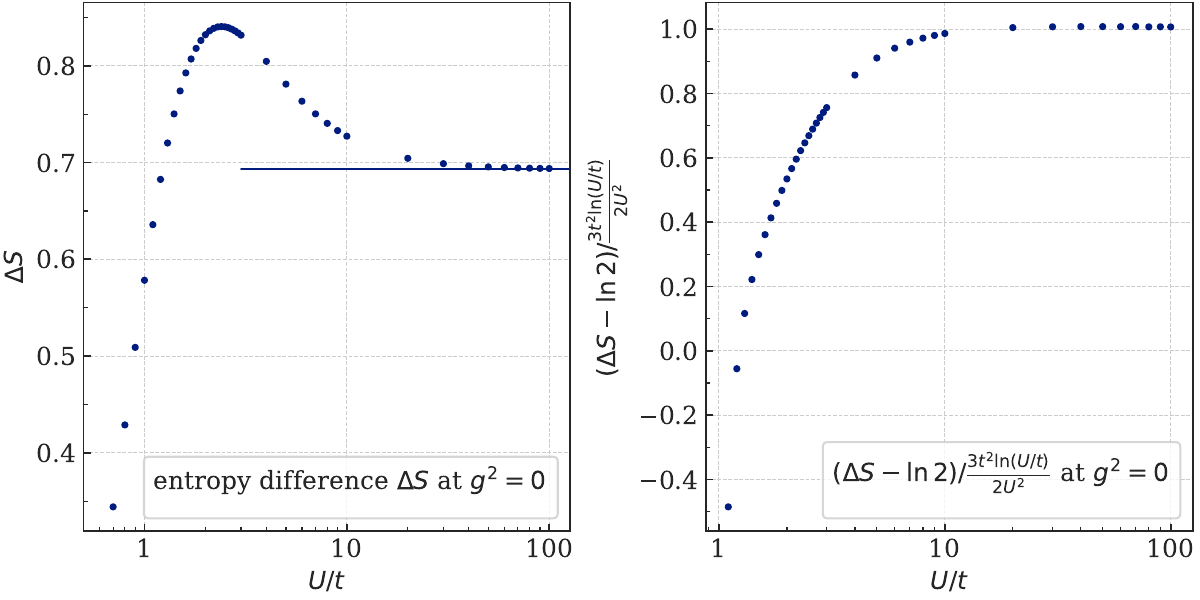}
  \caption{$\Delta S = S(\ell~\text{odd}) - S(\ell~\text{even})$ between odd and even $\ell$ for $N=2$ and $L=200$, where we choose $S(\ell~\text{odd}) = (S(99)+S(101))/2$ and $S(\ell~\text{even}) = S(100)$. In the left panel, we show $\Delta S$ as a function of $U/t$. We clearly see that as $U/t$ increases, $\Delta S$ approaches $\ln 2$, which is highlighted by the horizontal line. In the right panel, we show that the ratio $(\Delta S - \ln2)\big/ (\frac{3t^2}{2U^2}\ln\frac{U}{t})$ approaches $1$ as $U/t$ increases, which is predicted by \cref{eq:dee}.}
  \label{fig:entropy-difference}
\end{figure}

One way to test the above picture is to measure the entanglement entropy (EE) $S(\ell)$, between a subsystem of size $\ell$, i.e., all the lattice sites from $i=1$ to $i=\ell$ and the links between these sites, and the rest with size $L-\ell$ with open boundary conditions, where $L$ is even. With open boundary conditions, the ground state is no longer doubly degenerate, because the translated state of \cref{eq:SUN-dimer} would have unpaired fermions on the boundary and such a system will not obey the Gauss's law. Since \cref{eq:SUN-dimer} is a product state, $S(\ell)$ arises from a single dimer. When $\ell$ is odd the subsystem cuts a dimer and we get $S(\ell) = \ln \binom{N}{N/2}$; when $\ell$ is even, no dimers are cut and $S(\ell) = 0$. In the left panel of \cref{fig:entropy-difference}, we plot the difference $\Delta S := S(\ell~\text{odd}) - S(\ell~\text{even})$ as a function of $U/t$ for $N=2$ using tensor network methods as explained in \cref{app:ITensor}. We find that as $U/t \rightarrow \infty$, $\Delta S \rightarrow \ln 2$ as expected.

When $U$ is large but finite, the ground state is no longer a tensor product state of dimers, but a superposition of \cref{eq:SUN-dimer} with other states. This can be understood quantitatively by treating the hopping term $t$ as a perturbation, and in \cref{app:dimer}, we perform a $\frac{t}{U}$ expansion around the dimer state for arbitrary even $N$. In the case of $N=2$, we find that when $U/t \rightarrow \infty$, we have
\begin{align}
  S(\ell) =
  \begin{cases}
    \ln 2 + \frac{2t^2}{U^2}\ln\frac{U}{t} + O\big((\frac{t}{U})^2\big), & \ell \text{ is odd}, \\
    \frac{t^2}{2U^2}\ln\frac{U}{t} + O\big((\frac{t}{U})^2\big), & \ell \text{ is even}.
  \end{cases}
\end{align}
Hence   
\begin{align}
  \lim_{U/t\rightarrow \infty} \Delta S = \ln 2 + \frac{3t^2}{2U^2}\ln\frac{U}{t} + O\big((\frac{t}{U})^2\big).
  \label{eq:dee}
\end{align}
In the right panel of \cref{fig:entropy-difference}, we plot the ratio between $\Delta S - \ln 2$ and $\frac{3\ln (U/t)}{2U^2}$, which indeed approaches to 1 as $U/t \rightarrow \infty$.

Let us now consider how things change when $N$ is odd. In this case, we have $n_i = \frac{N\pm 1}{2}$. When $g^2=0$, all links have the same energy, and the two types of filling can be scattered everywhere in the lattice and every such state has the same energy. Thus, the ground states are exponentially degenerate in the system size when $t=0$. Taking into account the leading corrections due to the hopping term, the degeneracy will be lifted and the physics in this degenerate subspace can be described by one fermion hopping on the lattice, and the system is gapless. It is well-known that the ground state of one free fermion appears at half-filling, i.e., half of $n_i = \frac{N- 1}{2}$, and half of $n_i = \frac{N+ 1}{2}$. This implies that the ground state is automatically at half-filling when taking into account the hopping term.

On the other hand, when $g^2>0$, the links prefer to be in alternating $k=0$ and $k=\frac{N\pm 1}{2}$. Suppose $k_{2i}=0$ and $k_{2i-1} = \frac{N\pm 1}{2}$, and then we have $n_{2i-1} = \frac{N\pm 1}{2}$ and $n_{2i} = \frac{N\mp 1}{2}$. Each odd link has two choices, and thus the ground states can be viewed as direct products of directed dimer states. While the degeneracy of the ground state grows exponentially in the system size and the hopping term can break the degeneracy as in earlier gapless cases, here the fermion can only hop back and forth within a dimer. Therefore the fermions are extremely localized and the system is gapped. Again, the ground states are all at half-filling.

\subsubsection{Extension to the traditional model}\label{subsec:strong-traditional}
Many of the results we presented in this section also hold for the traditional lattice model, this is because as we discussed in \cref{subsec:qubit}, when $g^2 \gg t$, the qubit model has the same physics as the traditional model. The higher representations are suppressed and do not play an important role. In particular, when $g^2 \gg |U| \gg t$, the traditional model is also in the spin-chain phase, and when $U \gg g^2 \gg t$, it is also in the dimer phase.

In addition, when $-U/t \gg 1$ and for arbitrary $g^2$, the traditional model has the same physics as the qubit model and is in the spin-chain phase, because in this limit the gauge field and fermions are still decoupled. The only difference is that there are more constant gauge field configurations at $g^2=0$. All this implies that for all the regions beyond the shaded part in \cref{fig:phase-diagram-traditional}, we expect the qubit model to reproduce the physics of the traditional model.

On the other hand, when $U/t \gg 1$, the above analysis fails at $g^2 = 0$, where the traditional model becomes a free fermion with a generalized Hubbard interaction. As we discussed in \cref{subsec:full-Hamiltonian}, in this region the traditional model is gapless, and is described by the $\SU(2)_s$ WZW model when $N=2$. In contrast, the qubit model remains gapped. This discrepancy between the traditional model and qubit model might extend to small $g^2$, which is denoted by the shaded region in \cref{fig:phase-diagram-traditional,fig:phase-diagram-strong}.

\subsection{Confinement analysis}

Gauge theories in 2d are linearly confined. This means if we place a test quark that lives in the $k$-box single-column representation and its antiquark on lattice sites separated by a distance $r$, the energy $E_k$ of the system will increase linearly with $r$. This property is usually described by a confining string between the quarks, and the string tension is defined as
\begin{align}
  T_k = \frac{\d E_k}{\d r} .
\end{align}
However, when $r$ exceeds a certain length $r_0$, it will be energetically favorable to produce pairs of dynamical quarks to screen the string, i.e., the string breaks. In this section, we will investigate the string tension and compute the scale $r_0$ in the strong coupling limit with open boundary conditions. The ground state structures are different between the spin-chain phase and dimer phase leading to distinct confinement properties.

\subsubsection{Spin-chain phase}

Recall that the qubit model is in the spin chain phase when $-U/t\gg 1$ or $g^2/t \gg 1$. The fermions are in states $n_i = 0$ or $N$, and the links are constant $k_i = 0$, due to the open boundary condition. Let us now introduce a pair of test quark and anti-quark with $N$-ality $k$ and $-k$ in the system. Suppose no fermions of the system screen the test quarks, i.e., the string does not break, then the string tension is equal to the electric field term,
\begin{align}\label{eq:string-tension}
  T_k = g^2\frac{N+1}{2N}k(N-k),
\end{align}
On the other hand, it costs energy $2|U|k(N-k)$ to create $N-k$ pairs of fermions to screen the test quarks. Therefore the string will break when $T_k r_0 \sim 2|U|k(N-k)$, i.e.,
\begin{align}
\label{eq:breaking-scale}
r_0 \sim \frac{4N}{N+1} \frac{|U|}{g^2}.
\end{align}
When $r>r_0$, $E$ reaches a constant $E_0 \sim 2|U|k(N-k)$, which no longer depends on $r$, and the test quarks can move freely, because each of them is paired with a new dynamical quark to form a color singlet. One may also consider the case where the string is partially broken by $k'$ pairs of fermions, but it can be checked that these states always have higher energy than an unbroken string or a fully broken string.

\begin{figure}[htb]
\centering  \includegraphics[width=0.8\linewidth]{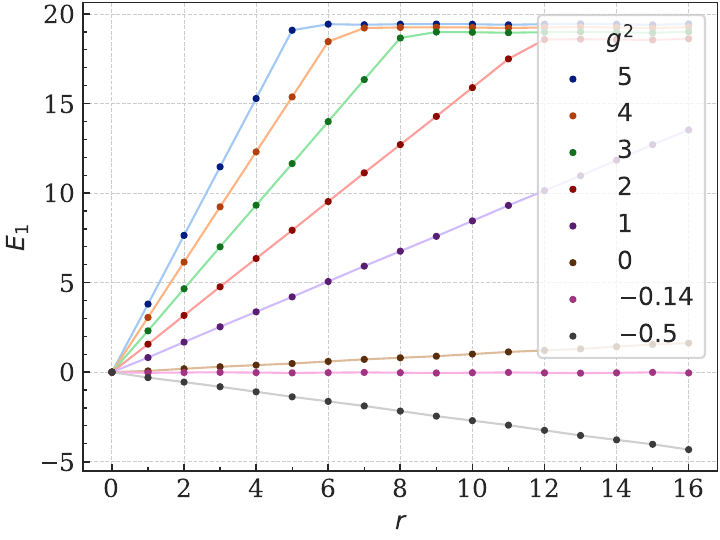}
\caption{Energy $E_1$ as a function of the distance $r$ between the test quarks at $N=2$, $k=1$ and $L=20$ for $U=-10$ and various $g^2$. $E$ grows linearly before the string breaks ($r<r_0$), and reach a constant $E_0 \approx 20$ after the string breaks ($r>r_0$).  As $g^2$ increases, the string tension increases, while $r_0$ decreases, and its value agrees with \cref{eq:breaking-scale}. The string tension is zero around $g^2 = -0.14$.}
  \label{fig:confinement}
\end{figure}

We have verified the above picture of the string formation and breaking numerically by measuring the energy $E_k$ as a function of $r$ using the tensor network method. A test quark on site $i$ is implemented by modifying Gauss's law on that site, i.e. instead of $\sum_a (G_i^a)^2 |\psi\> = 0$, we use $\sum_a (G_i^a)^2 |\psi\> = \frac{N+1}{2N}k(N-k)$. In \cref{fig:confinement}, we show our numerical result for the energy $E_1$ as a function of $r$ in the case of $N=2$ at $U=-10$ and various values of $g^2$ including two negative values of $g^2$. We see that before the string breaks, $E_1$ grows linearly with $r$, and the test quarks are linearly confined; after the string breaks, $E_1$ saturates to a constant $E_0 \approx 2|U|k(N-k) = 20$ as predicted above. This constant still shows some mild dependence on $g^2$, which is expected to disappear when $-U/g^2 \gg 1$. The string breaking scale $r_0$ also agrees with $\frac{80}{3g^2}$ from \cref{eq:breaking-scale}. The string tension vanishes at $g^2 = -0.14$, which is allowed for the qubit model as discussed at the end of \cref{subsec:qubit}.

\begin{figure}[htb]
\centering
\includegraphics[width=0.8\linewidth]{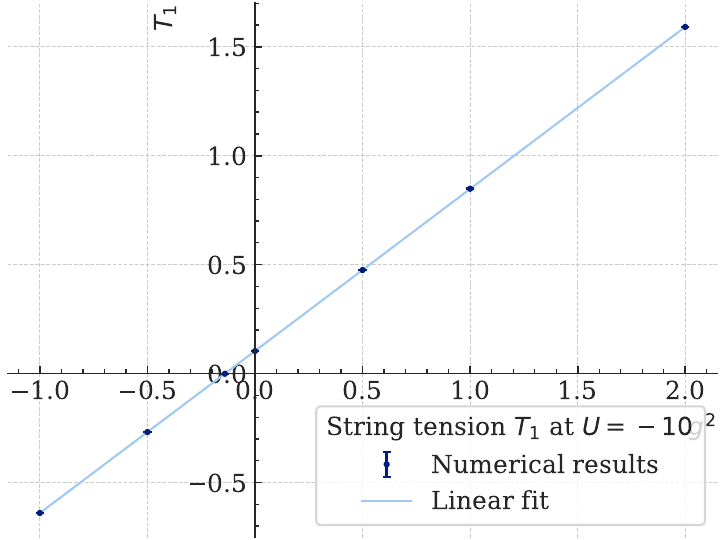}
\caption{The string tension $T_1$ as a function of $g^2$ at $U=-10$. The straight line shows a linear fit: $T_1 = 0.7438(2) g^2 + 0.1036(2)$.}
\label{fig:string-tension-g}
\end{figure}

Next, we verify \cref{eq:string-tension} by plotting the string tension $T_1$ as a function of $g^2$ in \cref{fig:string-tension-g}. In the case of $N=2$ and $k=1$, \cref{eq:string-tension} predicts that $T_1$ will be a linear function of $g^2$ with a slope of $0.75$. In \cref{fig:string-tension-g}, we find that our numerical results indeed fit to a line with a 
slope of $0.7438(2)$, close to the predicted value. However, we see that there is also a constant shift of $0.1036(2)$, which means $T_1>0$ when $g^2 = 0$. This is different from the traditional lattice model, where at $g^2 = 0$, the gauge field can be removed from the model and the theory is deconfined (see the discussion in \cref{subsec:KS-Hamiltonian}). In contrast, in the qubit model, the gauge field cannot be removed at $g^2 = 0$ due to the non-unitarity of the $U^{\alpha\beta}_{ij}$ operators. Therefore the model can be confining, and as we see here, it indeed confines. In \cref{app:generation}, we give a theoretical argument for this phenomenon, by showing that even when we set $g^2=0$, the $g^2$ term can be generated by the fermion hopping term in the sense of RG. 

\subsubsection{Dimer phase}

Let us now consider the confinement properties in the dimer phase, where $U/g^2 \gg 1$. In this limit, we discussed earlier that $n_i = \frac{N}{2}$ for $N$ even and $n_i = \frac{N\pm 1}{2}$ for $N$ odd. In both cases the links alternate between $k=0$ and $k=\frac{N}{2}$ (or $k=\frac{N\pm 1}{2}$ when $N$ is odd). The presence of test quarks and anti-quarks will shift the $N$-ality of the links in the middle by $k$. For simplicity, we assume the distance between the test quarks $r$ is even. Then the string tension is
\begin{align}
  T_k &= g^2\frac{N+1}{4N}\big(k(N-k) + (\frac{N}{2}+k) (\frac{N}{2}-k) - \frac{N^2}{4} \big) \nonumber\\
      &= g^2\frac{N+1}{4N}k(N-2k) 
\end{align}
for even $N$, and 
\begin{align}
  T_k &= g^2\frac{N+1}{4N}\big(k(N-k) + (\frac{N+1}{2}+k) (\frac{N-1}{2}-k) \nonumber\\
      &\hspace{4cm} - \frac{N+1}{2} \frac{N-1}{2} \big) \nonumber\\
      &= g^2\frac{N+1}{4N} k(N-2k-1) \quad 
\end{align}
for odd $N$. Note that when $N$ is odd, we have used the freedom of the directed dimer, i.e., $k=\frac{N+1}{2}$ to minimize the string tension. We can combine both these two cases into one equation by writing
\begin{align}\label{eq:string-tension-dimer}
  T_k &= g^2\frac{N+1}{2N}k(\lfloor\frac{N}{2}\rfloor-k) 
\end{align}
where $\lfloor x\rfloor$ is the floor function, which returns the greatest integer less than or equal to $x$, and we assume $k \leq \frac{N}{2}$. When $k > \frac{N}{2}$, $T_k$ can be obtained using the charge conjugation symmetry $k \mapsto N-k$.

The string is fully broken when the test quarks are screened by a pair of fermions with $N$-ality $-k'$ and $k'$, such that $k-k'=0$ or $k+k'= \lfloor\frac{N}{2}\rfloor$, which costs energy $2U{k'}^2$ for both even $N$ and odd $N$. Therefore the string breaks at
\begin{align}
  r_0 \sim \frac{4N}{N+1} \frac{U}{g^2} \frac{{k'}^2}{k(\lfloor\frac{N}{2}\rfloor-k)},
\end{align}
where
\begin{align}
  k' = \min\{k, \lfloor\frac{N}{2}\rfloor -k\}.
\end{align}

Notice that from \cref{eq:string-tension-dimer}, we see that the test quarks are confined unless $k=0$, $k = \lfloor\frac{N}{2}\rfloor$ and their charge conjugations. This implies that when $N=2,3$, the test quarks are always deconfined in the dimer phase, in contrast to the spin-chain phase discussed earlier.

\section{Critical Physics}
\label{sec:critical}
The strong coupling analysis performed in the previous section only teaches us the physics deep inside a phase where correlation lengths are small compared to the lattice spacing and far from the continuum physics we are interested in, especially in a massive phase. However, we must note that in the critical phase, one can still learn the physics of the WZW fixed points that emerge in the IR. 

In this section, we will focus on the physics near phase transitions. In particular, we will determine the phase boundary based on the continuum physics discussed in \cref{sec:continuum}, and show how the fixed points in \cref{fig:flow-diagram} are reproduced. In this section, we will focus on $N=2$.

\subsection{Critical points of the qubit model}
In this subsection, we explain how to determine the critical points of the qubit model precisely based on the continuum physics that describes it, as discussed in \cref{sec:continuum}. Recall that for $N=2$, we argued that the gapped and the gapless phases can be described by a perturbation of the WZW fixed point with current-current couplings, i.e.,
\begin{align}\label{eq:Lagrangian-N=2}
  \mathcal{L} = \mathcal{L}_\mathrm{WZW} + \lambda_c J_{cL}^a J_{cR}^a,
\end{align}
where $\mathcal{L}_\mathrm{WZW}$ is the Lagrangian of $\SU(2)_1$ WZW model that emerges in the IR. For a fixed $g^2$, the coupling $\lambda_c$ can be tuned by the Hubbard coupling $U$ \cite{Liu:2020ygc}, and the critical point $U=U_c$ corresponds to the $\lambda_c=0$ line in the continuum flow diagram of \cref{fig:flow-diagram}.

$U_c$ can be determined using the following argument in the continuum. At the critical point ($\lambda_c=0$), the lattice model has an emergent chiral symmetry $\SU(2)_{L} \times \SU(2)_{R}$. Therefore all energy eigenstates can be labeled by the corresponding quantum numbers $(s_L, s_R)$. On the other hand, away from the critical point, we have $\lambda_c \neq 0$, and the chiral symmetry $\SU(2)_L \times \SU(2)_R$ is broken down to the diagonal $\SU(2)$ symmetry. Thus for a generic Hubbard coupling $U$, the only good spin quantum number is $s_\mathrm{tot} = s_L + s_R$. In the $\SU(2)_1$ WZW model, it is well known that the lowest 5 states have $(s_L, s_R) = (0,0)$ and $(\frac{1}{2},\frac{1}{2})$, with momentum $k=0$ and $k=\pi$ respectively \cite{Affleck:1988px}, corresponding to the 5 primary operators in the theory. However, when $\lambda_c \neq 0$, the four states in $(\frac{1}{2},\frac{1}{2})$ are broken into a singlet and triplet, i.e.,    
\begin{align}
  (s_L, s_R) = (\frac{1}{2},\frac{1}{2}) \longrightarrow s_\mathrm{tot} = 1, 0.
\end{align}
Thus at generic values of $U$, the lowest five energy eigenstates of our qubit model consist of two singlets and one tiplet. At the critical point $U_c$ the singlets at $k=\pi$ will cross the triplet states, which can be used to determine $U_c$. Furthermore, it can be shown that
\begin{align}
  \<&J_L\cdot J_R\> = \frac{1}{2}\<(J_L+J_R)^2 - J_L^2 - J_R^2\> \nonumber\\
  &= \frac{1}{2}\big(s_\mathrm{tot}(s_\mathrm{tot}+1) - s_L(s_L+1) - s_R(s_R+1)).
\end{align}
Therefore, when $\lambda_c>0$, i.e., in the gapped phase, the triplet states have higher energy than the singlet state, and this ordering is flipped when $\lambda_c<0$.

\begin{figure}[htb]
  \centering
  \includegraphics[width=0.8\linewidth]{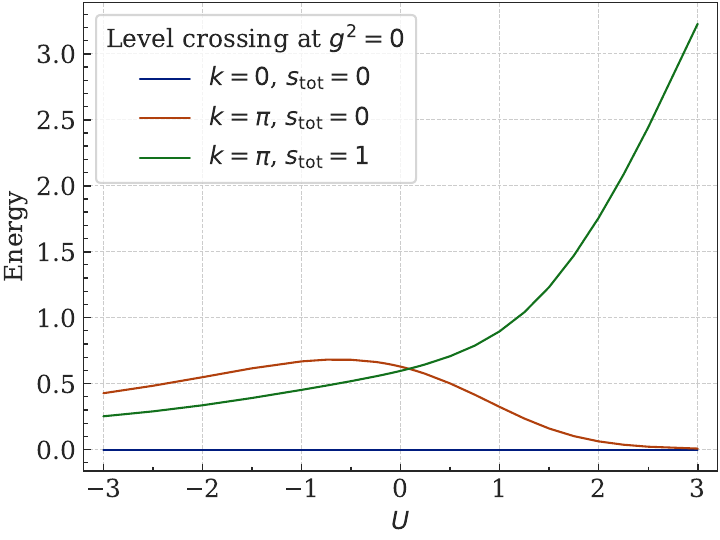}
  \caption{Level crossing of the lowest five energy eigenstates as a function of $U$ at $g^2=0$ and $L=10$ with periodic boundary conditions. At the crossing point $U_c(L=10) =0.08534(7)$, the four $k=\pi$ states become degenerate due to the vanishing of the marginal coupling $\lambda_c$ and emergent chiral symmetry $\SU(2)_L \times \SU(2)_R$. When $U>U_c$, the triplet states have higher energy than the singlet state, indicating $\lambda_c>0$ and the system is gapped. Besides, when $U$ becomes larger, the two $s_\mathrm{tot}=0$ states with $k=0, \pi$ quickly become degenerate. These results agree with our strong coupling analysis.}
  \label{fig:level-crossing}
\end{figure}

We computed the lowest five energy eigenvalues in our model using the tensor network method. In \cref{fig:level-crossing}, we plot these five eigenvalues as a function of $U$ at $g^2=0$ and $L=10$ with periodic boundary conditions. 
We see that when $U>U_c$, the triplet states have higher energy than the singlet state, indicating $\lambda_c>0$ and the system is in the gapped phase. As $U$ becomes larger, the two $s_\mathrm{tot}=0$ states with $k=0, \pi$ quickly become degenerate as expected in the dimer phase. These results agree with our strong coupling analysis in \cref{subsec:phase}: when $U/t \gg 1$, the system is gapped, and the ground state is dimerized with double degeneracy for $N=2$.

\begin{figure}[htb]
  \centering
  \includegraphics[width=0.8\linewidth]{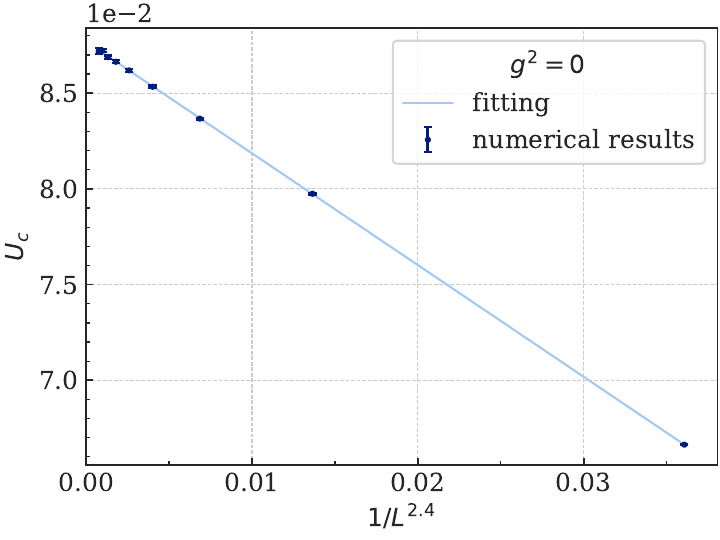}
  \caption{Extrapolation of the critical point $U_c$ as a function of $L$ at $g^2=0$. $L$ ranges from $4$ to $20$. The fitting to \cref{eq:UcL} given by $U_c(L) = 0.08769(3) - 0.583(9)L^{-2.39(1)}$, which means when $L\rightarrow \infty$, $U_c \approx 0.08769(3)$.}
  \label{fig:critical-point-extrapolation-g=0}
\end{figure}

Of course, there might be other irrelevant operators contained in $U$ which will shift the location of $U_c$, making it a function of the system size $L$. However, we expect these irrelevant operators to disappear as $L \rightarrow \infty$. In order to determine the true critical point we perform an extrapolation of $U_c(L)$ using data from $L = 4$ to $L = 20$. Empirically we find that $U_c(L)$ satisfies the relation
\begin{align}\label{eq:UcL}
  U_c(L) = U_c(\infty) + AL^{-a}.
\end{align}
To show this, in \cref{fig:critical-point-extrapolation-g=0}, we plot $U_c$ as a function of $L^{-a}$, where we have fixed $a=2.4$ based on our fits, and we see that $U_c(L)$ falls perfectly on a straight line. The exponent $a$ is expected to be related to the dimension of the most relevant term besides the marginal term $\lambda_c$ contained in the Hubbard coupling. By performing an extrapolation in $L\rightarrow\infty$, we can determine $U_c=0.08769(3)$ quite precisely at $g^2=0$.

\begin{figure}[htb]
  \centering
  \includegraphics[width=0.8\linewidth]{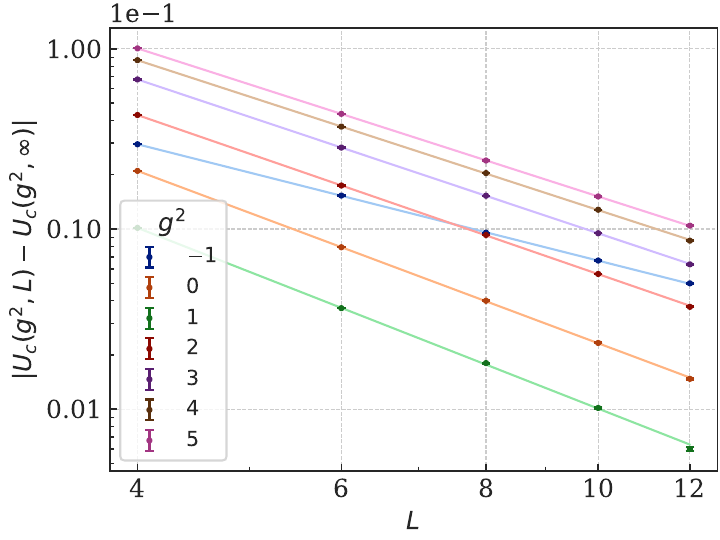}
  \caption{Extrapolation of the critical point $U_c$ as a function of $L$ for various $g^2$ in log-log scale. The fact that for any fixed $g^2$ the data fit to a straight line indicates that $U_c(g^2, \infty)$ satisfies \cref{eq:UcL}.}
  \label{fig:critical-point-extrapolation-g}
\end{figure}

We repeated the above analysis for different values of $g^2$ and found that \cref{eq:UcL} continues to be valid if we allow the fit parameters $A$ and $a$ to be $g^2$-dependent. In \cref{fig:critical-point-extrapolation-g} we plot $|U_c(g^2, L) - U_c(g^2, \infty)|$ against $L$ on a log-log plot, and observe that for each fixed value of $g^2$, our data fall on a straight line consistent with \cref{eq:UcL}. The original data and the extrapolated $U_c(g^2, \infty)$ are tabulated in \cref{tab:Uc} in \cref{app:data}.

\begin{figure}[htb]
  \centering
  \includegraphics[width=0.8\linewidth]{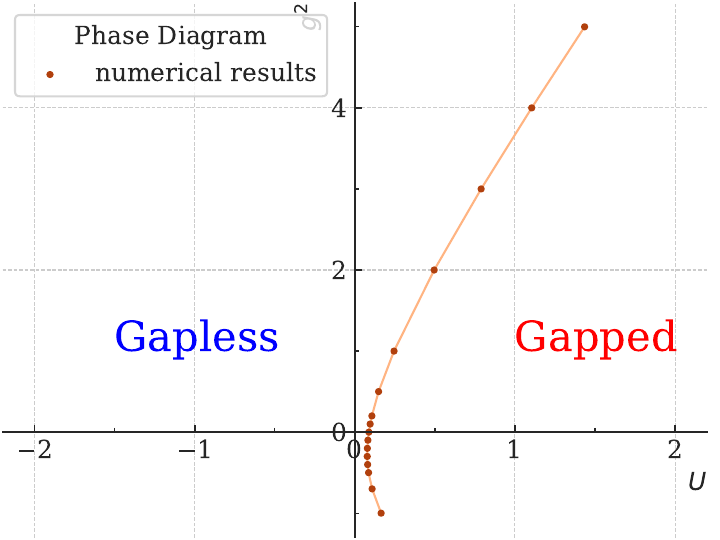}
  \caption{Phase diagram of the qubit model \cref{eq:ks-with-Hubbard} at $N=2$. The orange line is the phase boundary which separates a gapped phase with doubly degenerate ground states, and a gapless phase.}
  \label{fig:phase-diagram}
\end{figure}

In \cref{fig:phase-diagram}, we plot the phase diagram of the qubit model in the $U$ and $g^2$ plane using the values of $U_c(g^2, \infty)$ in \cref{tab:Uc}. The orange line is the phase boundary between the gapped and the gapless phase. When $g^2>0$, we notice that $U_c(g^2)$ increases as $g^2$ increase. This is consistent with our result from strong coupling expansion: when $-U/t\gg 1$ or $g^2/t\gg 1$, we found that the two couplings $U$ and $g^2$ always appear in the combination $\frac{N+1}{2N} g^2 - 2U$ in the various relations we derived. Therefore we expect that certain physics only depends on this linear combination. This suggests that the phase boundary may also appear according to this relationship. In fact, we can check that when $g^2$ is large, the phase boundary in \cref{fig:phase-diagram} is linear with a slope
\begin{align}
  \frac{\Delta U}{\Delta g^2} \approx 0.33,
\end{align}
while the slope of the line obtained when $\frac{N+1}{2N} g^2 - 2U$ remains a constant is 
\begin{align}
  \frac{\d U}{\d g^2} = \frac{N+1}{4N},
\end{align}
which is $0.375$ for $N=2$.

The phase diagram of the qubit model is similar to that of the traditional lattice model shown in \cref{fig:phase-diagram-traditional}, except that the phase boundary does not reach $g^2 = U = 0$. This difference arises because when $g^2=0$ the fermions are not free, unlike the traditional model. As we explain in \cref{app:generation}, from an RG perspective, when we begin with $g^2=0$ in the qubit model, an effective $g^2$ term is generated by the hopping term, which does not happen in the traditional model. Hence, in the qubit model, even when $g^2=0$ fermions interact and an effective marginally irrelevant coupling $\lambda_c$ (see \cref{fig:flow-diagram}) is generated, which needs to be canceled by a positive $U$.

One of the goals of qubit regularization is to recover the continuum physics discussed in \cref{sec:continuum}. That physics is described by two fixed points, the free-fermion fixed point and the WZW fixed points and the flow from one to the other. This physics can be recovered in the traditional lattice model \cref{eq:ks-with-Hubbard}, by tuning to $g^2=U=0$. How well can we recover this physics of the two fixed points using the qubit model? Based on the above arguments it seems it will be difficult to reach free-fermion fixed point, but how close are we to it? On the other hand, we may be able to reach points close to the vertical line ($\lambda_c=0$, $\tilde{g}^2 > 0$) in the flow diagram of \cref{fig:flow-diagram} by tuning our qubit model to the critical points $U_c$. In fact, we have already argued that the IR physics of the qubit model in the gapless phase will be described by the same continuum field theory as the IR physics of the traditional model. This is the physics of the WZW fixed point.

In the following sections, we will explore how well our qubit model is able to recover the expected universal behavior close to the two fixed points. 

\subsection{The WZW fixed point}

There are several universal properties of a fixed point of an RG flow, and in 2d, the central charge is an important one. If our qubit model in the gapless phase flows to the WZW fixed point in the IR, then the central charge of our lattice model in the IR should be $c_{\rm WZW}=1$.

In a gapless phase in $1+1$d, there is a simple way to compute the IR central charge using EE $S(\ell)$ defined in \cref{subsec:dimer}, except that we now assume periodic boundary conditions. In an infinite system, we have \cite{Holzhey:1994we}
\begin{align}
  S(\ell) = \frac{c_{\IR}}{3} \ln\frac{\ell}{a} + \mathrm{const.} ,
  \label{eq:eeinf}
\end{align}
where we assume the subsystem of size $\ell$ shares two boundaries (points) with its environment. Here $c_{\IR}$ is the central charge of the gapless system. If $L$ is finite, we can use an exponential map from the cylinder to a plane, and one finds that $\ell$ is replaced by $\frac{L}{\pi a} \sin\frac{\pi\ell}{L}$ \cite{Calabrese:2004eu,Calabrese:2009qy}, i.e.,
\begin{align}\label{eq:cIR}
  S(\ell) = \frac{c_{\IR}}{3} \ln\Big(\frac{L}{\pi a} \sin\frac{\pi\ell}{L}\Big) + \mathrm{const.}
\end{align}
By measuring $S(\ell)$ as a function of $\ell$ and fitting to \cref{eq:cIR} we can obtain the central charge. As explained in \cref{app:central-charge}, the central charge obtained in this way corresponds to the central charge of the IR fixed point, and hence is denoted as $c_\IR$.

\begin{figure}[htb]
  \centering
  \includegraphics[width=0.8\linewidth]{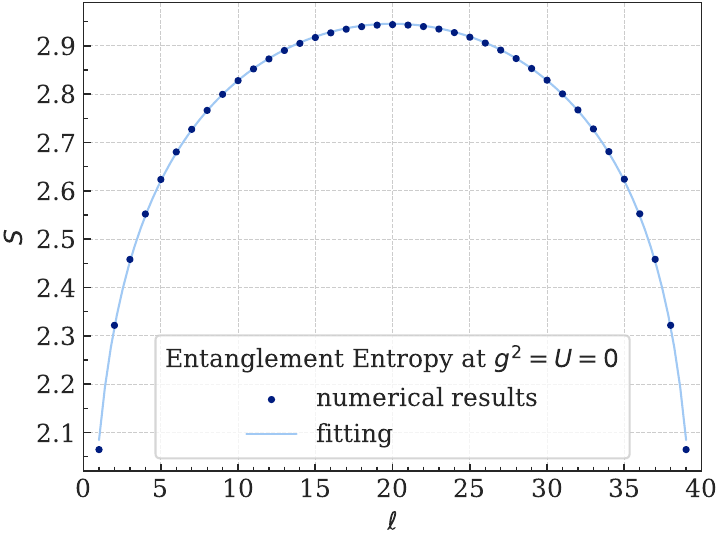}
  \caption{The EE $S$ at $g^2 = U = 0$ for $L=40$. The curve is fitting to \cref{eq:cIR}, which gives $c_\IR=1.015(3)$.}
  \label{fig:EE-N=40}
\end{figure}

In \cref{fig:EE-N=40}, we measure $S$ between two subsystems of size $\ell$ and $L-\ell$, in the ground state of the qubit model at $g^2 = U = 0$ with periodic boundary conditions. By fitting to \cref{eq:cIR}, we find that $c_\IR = 1.015(3)$, which is close to the expected result of $c_\wzw=1$.

\begin{figure}[htb]
  \centering
  \includegraphics[width=0.8\linewidth]{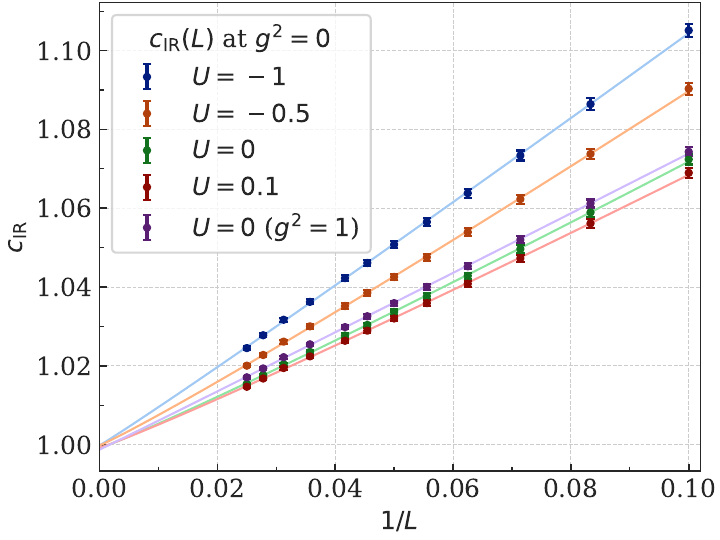}
  \caption{Extrapolation of the central charge $c_\IR$ as a function of $L$ at various parameters in the massless phase. The central charge $c_\IR$ ranges from $0.9988(7)$ to $0.9998(9)$, strongly supporting our theoretical analysis. For display purposes, the error bars in this plot are $1/10$ of the true error bar obtained from fitting to \cref{eq:cIR}.}
  \label{fig:central-charge-extrapolation}
\end{figure}

The value of $c_\IR$ can be improved by noting that \cref{eq:cIR} is a result of conformal field theory that assumes that there are only two scales in the problem. However, in the lattice theory there are several other scales, making the central charges obtained in this way depend on the system size $L$ and will be denoted as $c_\IR(L)$. This is because as explained in \cref{app:central-charge}, any mode longer than $L$ will contribute to $c_\IR$, making it larger than what it would be. In \cref{fig:central-charge-extrapolation}, we plot $c_\IR(L)$ against $1/L$ for various $U$ and $g^2$. We again fit our numerical results to a form similar to \cref{eq:UcL},
\begin{align}
  c_\IR(L) = c_\IR + AL^{-a}.
\end{align}
For the couplings used in \cref{fig:central-charge-extrapolation}, $c_\IR$ ranges from $0.9988(7)$ to $0.9998(9)$, which is very close to the theoretical value $c_{\rm WZW}=1$. This provides strong evidence that our qubit model indeed flows to the WZW fixed point. 

\subsection{The free-fermion fixed point}\label{subsec:free-fermion}

One of the main challenges for the qubit model is to recover the physics of the free-fermion fixed point with central charge $c_\fr=2$.
Notice that if the qubit model can be tuned to a point where $\tilde g^2 = 0$ and $\lambda_s, \lambda_c \leq 0$ in \cref{eq:Lagrangian}, then the free-fermion fixed point can be realized as an IR fixed point. This of course requires fine-tuning, and it is possible that our model with two parameters $g^2$ and $U$ might not be sufficient to achieve this since we need at least three parameters $\tilde{g}^2$, $\lambda_s$ and $\lambda_c$ to achieve this in the continuum. It is also possible that the link Hilbert space is too small to mimic the continuum gauge coupling $\tilde{g}^2$. Still, we can explore if there are UV fixed points in our lattice model and what is the central charge at those fixed points.

We can again measure this central charge using EE, but from a different perspective. For gapped systems, $S(\ell)$ satisfies the area law, and in $1+1$d, when the subsystem size $\ell$ is much larger than the length scale $\xi$, it saturates to a constant value
\begin{align}\label{eq:cUV}
  S = \frac{c_\UV}{6} \ln\frac{\xi}{a} + \mathrm{const.} ,
\end{align}
assuming we use open boundary conditions and there is only one boundary between the two subsystems. As explained in \cref{app:central-charge}, the central charge here detects the UV physics, because when we fix the length $\xi$ in physical units, $c_\UV$ characterizes how $S(\ell)$ diverges as the lattice spacing $a$ (UV scale) goes to zero when the couplings are tuned to a critical point.

\begin{figure}[htb]
  \centering
  \includegraphics[width=0.8\linewidth]{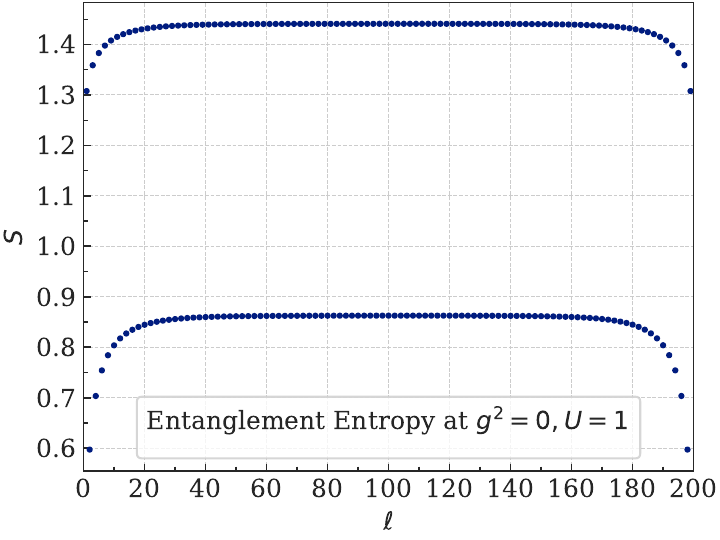}
  \caption{The plot $S(\ell)$ as a function of $\ell$ at $g^2 = 0$ and $U=1$ for $L=200$. We see that $S(\ell)$ saturates to different constants depending on whether $\ell$ is even or odd when $\ell \gg \xi$, which }
  \label{fig:EE-N=200}
\end{figure}

Let us first discuss how to compute $S$ from $S(\ell)$ and $\xi$. In \cref{fig:EE-N=200}, we plot $S(\ell)$ between two subsystems of size $\ell$ and $L-\ell$, in the ground state of our model at $g^2 = 0$ and $U = 1$ with open boundary conditions and $L=200$. Since we are in a gapped phase, we expect $S(\ell)$ to saturate. Interestingly, we observe that the saturation values of $S(\ell)$ oscillate between even and odd $\ell$. This oscillating behavior between even and odd $\ell$ for open boundary conditions has been studied earlier for 1d gapless systems \cite{Laflorencie:2005duh,Calabrese:2009us}, where it was found that the oscillations decay to zero from the boundary with a universal power law. However, in a gapped phase, $S(\ell)$ for both even and odd values of $\ell$ saturates to different values, and the difference is expected to approach zero only as getting closer to the critical point, as is shown in \cref{fig:entropy-difference}. In \cref{subsec:dimer}, we gave an explanation for the difference using the dimer ground state picture. When $\ell$ is odd, we argued that $S(\ell)$ contains an extremely local dimer contribution, which can be viewed as a lattice artifact. This suggests that when analyzing the emergent continuum theory, it is important to use the even $\ell$ results. For $g^2=0$ and $U=1$ we find that $S=0.8626$. $S$ at various $g^2$ and $U$ are tabulated in \cref{tab:S-xi}.

\begin{figure}[htb]
  \centering
  \includegraphics[width=0.8\linewidth]{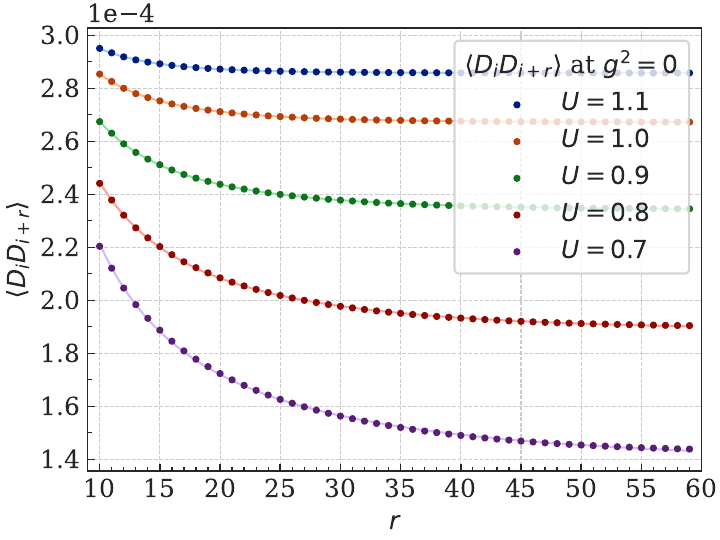}
  \caption{Dimer correlation function in the massive phase. Solid lines are fitting to \cref{eq:dimer-correlation}}
  \label{fig:dimer-correlation}
\end{figure}

In order to find $c_\UV$ we also need to determine the correlation lengths $\xi$ at the same couplings where we computed $S$. To determine $\xi$ we measured the correlation function of the dimer operator in the charge sector in the ground state, which is defined as
\begin{align}\label{eq:dimer}
  D_i &:= (-1)^i\frac{1}{2}(Q^z_iQ^z_{i+1}-Q^z_{i-1}Q^z_i),
\end{align}
where $Q^z_i = n_i - 1$. In the massive phase of $\SU(2)_1$ WZW model, the dimer correlation function is expected to decay as \cite{Liu:2020ygc}
\begin{align}\label{eq:dimer-correlation}
  \<D_iD_{i+r}\> = \frac{A}{r} \e^{-\frac{r}{\xi}},
\end{align}
where $A$ is a constant depending on $U$. In \cref{fig:dimer-correlation}, we plot the dimer correlation function at $g^2=0$ for various values of $U$ in the massive phase. By fitting to \cref{eq:dimer-correlation}, we extract the correlation length $\xi$ for each value of $U$. The extracted values of $\xi$ are tabulated in \cref{tab:S-xi}.

\begin{figure}[htb]
  \centering
  \includegraphics[width=0.8\linewidth]{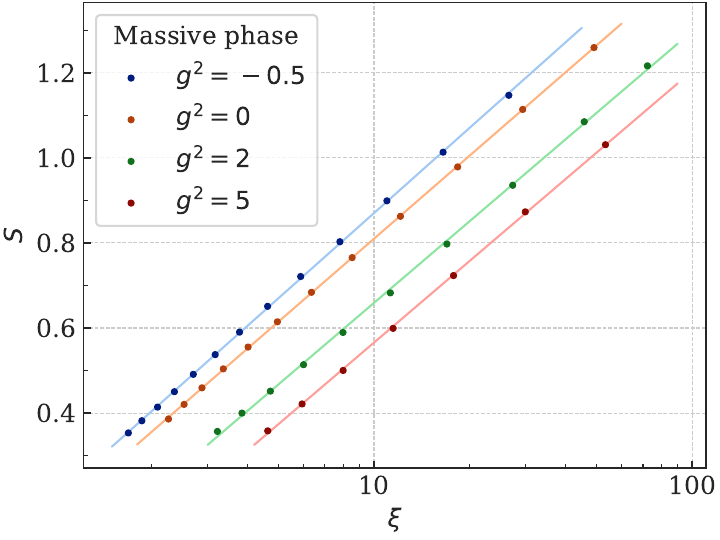}
  \caption{The EE $S$ at $\ell=100$ as a function of correlation length $\xi$. Each color shows a fixed $g^2$ and varying $U$, where the smallest $U$ for each $g^2$ are $0.8$, $0.7$, $1$, and $2$ from top to bottom, and decreasing step is $0.1$. The lines show fitting to \cref{eq:cUV}. From the fitting we extract $c_\UV$ to be $1.737(6)$, $1.693(4)$, $1.66(1)$, $1.66(1)$ from top to bottom.}
  \label{fig:entropy-xi-g}
\end{figure}

For several $g^2$ and $U$ values we determined both $\xi$ and $S$. In \cref{fig:entropy-xi-g}, we plot $S$ as a function of $\xi$ on a logarithmic scale, by varying $U$ for various fixed values of $g^2$. We find the data fall on straight lines in agreement with \cref{eq:cUV}. From the slopes we can determine $c_\UV$, and we find $c_\UV$ slightly increases as $g^2$ decreases. For example we obtain $c_\UV = 1.737(6)$ at $g^2=-0.5$.

Note that $c_\UV$ is between $c_\fr=2$ and $c_\wzw=1$, which can be understood as follows. As discussed in \cref{app:central-charge}, $c_\UV$ measures the degrees of freedom whose correlation lengths diverge in the lattice unit. At the free-fermion fixed point, both $\xi$ in the charge channel and $\xi_s$ in the spin channel diverge in the lattice unit as we get close to the critical point. Each of them contributes a central charge of $1$ to $c_\UV$, together making $c_\UV = 2$. On the other hand, near the WZW fixed point, the spin channel is gauged, and the correlation $\xi_s$ of the spin operators is highly confined due to the Wilson line attached between them. Therefore only $\xi/a$ in the charge channel diverges, and we have $c_\UV = 1$. However, at some intermediate coupling that is not close enough to the critical point, it is possible that both $\xi_s/a$ and $\xi/a$ grow but $\xi/\xi_s$ also grows. Thus, the spin channel only contributes partially to the growth of the EE, and we observe $1 < c_\UV < 2$.

In order to determine the true UV central charge, we need to get closer to the critical point. This will make $\xi$ much larger than the lattice spacing $a$, which in turn requires lattice sizes $L > \xi \gg a$ in order for the entanglement entropy to saturate. While this would be challenging numerically, we can speculate the following three possible results: 1. $c_\UV = 2$ at some fine-tuned region close to the free-fermion critical point; 2. $c_\UV = 1$ at all critical points because our model does not contain the free-fermion critical point; 3. $1 < c_\UV < 2$, indicating a new UV fixed point with central charge between $1$ and $2$, which would be quite exotic and interesting by itself.

If it turns out that our model does not contain the free-fermion critical point, we can still modify our model in several ways to look for it. For example, we could try to introduce new terms in the qubit model motivated by the RG flow. The fact that when $g^2$ becomes negative $c_\UV$ increases suggests that we may need to introduce unconventional electric field couplings. Another way to modify our model would be to increase the dimension of the link Hilbert space by including higher representations, as was done in \cite{Banuls:2017ena}. As we argued in \cref{subsec:dimer}, for the current link Hilbert space, there are only two gauge-invariant states in the spin sector for $N=2$. After including links carrying spin $1$, there will be an exponentially large number of gauge-invariant states. Furthermore, when measuring $\xi_s$ using correlation functions of the spin operators, which have spin $1$, we have to attach Wilson lines between them to form gauge-invariant observables. With spin $1$ links, this can be easily done using spin $1$ Wilson lines. However, if we only have spin $0$ and spin $1/2$ links as in our current qubit model, the spin $1$ Wilson line is not allowed in the theory, and we must use two spin $1/2$ Wilson lines, with one of them winding around the whole space, which is very unnatural. When $N \geq 3$, the spin $1$ links of $N=2$ would correspond to links in the adjoint representation of $\SU(N)$.

\section{Conclusions}
In this work, we explored the potential of using qubit regularization to study 2d massless QCD, i.e., $\SU(N)$ gauge theory with a single flavor of massless Dirac fermions in the fundamental representation. This continuum theory contains a free-fermion fixed point with central charge $N$ in the UV and a coset WZW fixed point with central charge $1$ in the IR. By adding current-current interactions to the continuum theory, which are usually generated on the lattice, we argued that the continuum theory without these interactions is at the boundary of two phases. We study this theory using a simple qubit model obtained by truncating the link Hilbert space of the traditional lattice theory to the single-column Young diagram representations, and introducing a generalized Hubbard coupling.

Using strong coupling analysis, we were able to show that our qubit model has a spin-chain phase when $-U/t \gg 1$ or $g^2/t \gg 1$, which is gapless for $N=2$, and a gapped dimer phase when $U/t \gg 1$. We also computed the confinement properties of these two phases using the string tension of test quarks as an observable. Using tensor network methods, which were limited to $N=2$, we found that the string tension is non-zero even at $g^2=0$, and it vanishes at a negative $g^2$. We gave a theoretical explanation of this by showing the electric field term is generated by the fermion hopping term in the qubit model. 

While some of the features of the qubit model are clearly different from the traditional model, the phase diagrams of the two theories are very similar. In fact, the universal aspects of the IR physics in the gapless phases match. Using the tensor network method in the $N=2$ case, we found that the low-lying spectrum can be understood as an $\SU(2)_1$ WZW model, where the phase transition is driven by a marginal current-current coupling. Using this understanding, we were able to determine the phase boundary of the qubit model very precisely. We also showed that the IR central charge in the gapless phase is $c_\wzw=1$ as expected in the continuum theory.

An important difference between the qubit model and the traditional model that remains unresolved is the location of the free-fermion fixed point. In the traditional model, this point is known to be located at $g^2=U=0$ on the phase boundary. In the qubit model, determining this point requires fine-tuning. At the free-fermion fixed point, we expect $c_\UV = c_\fr = N$, but our numerical results show $c_\UV \approx 1.7$ for $N=2$. We gave an explanation for this discrepancy based on our understanding of the physical meaning of $c_\UV$, and argued that in order to determine the true $c_\UV$, a much larger lattice will be needed, which is beyond the scope of the present study. We also discussed several possible ways to extend our qubit model to include the free-fermion critical point.

Our numerical work has been restricted to $N=2$, and it would be interesting to extend our work to higher values of $N$. As we explained in the introduction, while there is already work with $N=3$ \cite{Silvi:2019wnf,Rigobello:2023ype}, most of the previous work has not focused on understanding the RG flow and phase diagram of the qubit models and recovering the fixed point structure expected in the continuum. In this work, we have developed the framework to address this question in a systematic manner.

\acknowledgments

H.L. is grateful to M. C. Ba\~nuls, P. Emonts, S. K\"uhn, M. Nguyen, H. Singh, Y. Tanizaki, M. \"Unsal and X. Yao for helpful discussions. GPT-4 from OpenAI was used for this work. This work was carried out under the auspices of the National Nuclear Security Administration of the U.S. Department of Energy at Los Alamos National Laboratory under Contract No. 89233218CNA000001. H.L. and T.B. were supported by the Quantum Science Center, a National Quantum Science Initiative of the Department of Energy, managed by Oak Ridge National Laboratory. H.L. was also supported by the U.S. Department of Energy, Office of Science, Nuclear Physics program. T.B. was also supported by the U.S. Department of Energy, Office of Science—--High Energy Physics Contract KA2401032 to Los Alamos National Laboratory. S.C. was supported by a Duke subcontract of this grant and also supported in part by the U.S. Department of Energy, Office of Science, Nuclear Physics program under Award No. DE-FG02-05ER41368. This work was performed in part at Aspen Center for Physics, which is supported by National Science Foundation grant PHY-2210452.

\bibliographystyle{apsrev4-2} 
\showtitleinbib
\bibliography{refs,Refs2,Refs-QS}

\clearpage
\appendix

\section{Convention on Lie algebras}
\label{app:Lie}

Let $T_\lambda^a: a=1, \cdots, \dim(\mathfrak{g})$ be a basis of the Lie algebra $\mathfrak{g}$ in the representation $\lambda$. When $\lambda$ is omitted, $T^a$ is assumed to be in the fundamental representation, which is normalized as
\begin{align}
  \tr(T^aT^b) =
  \begin{cases}
    \delta^{ab} & \text{ for } \so(2N), \\
    \frac{1}{2}\delta^{ab} & \text{ for } \su(N).
  \end{cases}
\end{align}
In this way, we make sure that when $\su(N)$ is viewed as a subalgebra of $\so(2N)$, the corresponding elements are normalized in the same way.

The structure constant $f^{abc}$ is defined as
\begin{align}
  [T^a, T^b] = \i f^{abc} T^c,
\end{align}
which is totally antisymmetric. The \emph{adjoint representation} $T_\mathrm{adj}^a$ is defined as
\begin{align}
  (T_\mathrm{adj}^a)_{bc} := -\i f^{abc}.
\end{align}
The \emph{Killing form} on $\mathfrak{g}$ is a symmetric bilinear form defined as
\begin{align}
  g^{ab} := \tr(T_\mathrm{adj}^a T_\mathrm{adj}^b) = (-\i f^{acd}) (-\i f^{bdc}) = f^{acd}f^{bcd}
\end{align}
and it is positive definite for compact Lie groups. In particular, we have
\begin{align}
  g^{ab} =
  \begin{cases}
    (2N-2) \delta^{ab} & \text{ for } \so(2N), \\
    N \delta^{ab} & \text{ for } \su(N).
  \end{cases}
\end{align}

$g_{ab}$ is defined to be the inverse of $g^{ab}$, i.e., $g_{ab}g^{bc} = \delta_a^c$. The \emph{quadratic Casimir operator} $C_2(\lambda)$ for an irrep $\lambda$ is defined as
\begin{align}
  C_2(\lambda) := g_{ab} T_\lambda^a T_\lambda^b =: c_2(\lambda) \mathbbm{1}_\lambda ,
\end{align}
where we have used the fact that $C_2(\lambda)$ is always proportional to identity. Using Weyl's formula, one can calculate that for $k$-box single column representations of $\su(N)$,
\begin{align}
  c_2(k) = \frac{N+1}{2N^2}k(N-k).
\end{align}
Therefore the electric field operators 
\begin{align}
  L^{a2} &= R^{a2} = \delta_{ab}T_\lambda^a T_\lambda^b = N C_2 \nonumber\\
  &= \sum_{k=0}^{N-1} \frac{N+1}{2N}k(N-k)\mathbbm{1}_k
\end{align}
in our qubit regularization.

\section{Strong coupling expansions}
\subsection{The spin-chain phase}
\label{app:spin-chain}

In this appendix, we derive the leading order effect of the hopping term $t$ in the qubit model when $-U/t \gg 1$ or $g^2/t \gg 1$ as discussed in \cref{subsec:spin-chain}. When $t=0$, we found that the ground states are $2^L$-fold degenerate. The degeneracy of these states will be lifted by the fermion hopping term, which can be studied using degenerate perturbation theory.

Since the hopping term couples only nearest neighbor sites, we can focus on the Hamiltonian on two such sites with a single link,
\begin{align}
  H &= H_0 + t V,
\end{align}
where
\begin{align}
  H_0 &= \frac{g^2}{2} \big( L^{a2} + R^{a2} \big) - U \sum_{i=1,2}n_i(N-n_i), \nonumber\\
  V &= t(c^{\alpha\dagger} U^{\alpha\beta} d^\beta + \hc).
\end{align}
Using the notation for states introduced in \cref{subsec:spin-chain}, we have
\begin{align}
  &H_0 |\{k_1\}, \{n_1, n_2\}, \alpha\> = E_{\{k_1\}, \{n_1, n_2\}}|\{k_1\}, \{n_1, n_2\}, \alpha\>, \nonumber\\
  &E_{\{k_1\}, \{n_1, n_2\}} = \frac{N+1}{2N} g^2 k_1(N-k_1) - U\!\! \sum_{i=1,2}n_i(N-n_i).
\end{align}
The ground state subspace of $H_0$ is 4d, whose basis can be labeled by the fermion numbers as $|00\>$, $|0N\>$, $|N0\>$ and $|NN\>$. The gauge link is always $k_1=0$ and can be omitted.

The hopping term $V$ creates mixing between these states, and thus breaks the degeneracy. The effective Hamiltonian $H_\eff$ in the ground state subspace can be determined by calculating its matrix elements. Since 
\begin{align}
  V|00\> = V|NN\> = 0,
\end{align}
the matrix elements of $H_\eff$ involving these states are zero. Hence, $H_\eff$ only has four non-zero matrix elements. We only have to determine two of them: $\<0N|H_\eff|0N\>$ and $\<N0|H_\eff|0N\>$, because the other two can be obtained using particle-hole symmetry. The leading order of $\<0N|H_\eff|0N\>$ consists of one fermion hopping to the left and then hopping back,
\begin{align}
  &\quad \<0N|H_\eff|0N\> = t^2 \<0N|V\frac{1}{E_0-H_0}V|0N\> \nonumber\\
  &= t^2 \frac{\sum_{\alpha\beta}\<0N|U^{\alpha\beta\dagger} U^{\alpha\beta}|0N\>}{-(\frac{N+1}{2N} g^2 - 2U)(N-1)} \nonumber\\
  &= -\frac{N}{N-1} \frac{t^2}{\frac{N+1}{2N} g^2 - 2U} =: -2J_z.
\end{align}
The leading order of $\<N0|H_\eff|0N\>$ consists of every fermion on the right hopping to the left, 
\begin{align}
  &\quad \<N0|H_\eff|0N\> = t^N \<N0|V \Big(\frac{1}{E_0-H_0}V\Big)^{N-1} |0N\> \nonumber\\
  &= t^N \frac{ \<N0|V^N |0N\> }{(-1)^{N-1}(\frac{N+1}{2N} g^2 - 2U)^{N-1} ((N-1)!)^2}  \nonumber\\
  &= (-1)^{N-1} \frac{N}{(N-1)!} \frac{t^N }{(\frac{N+1}{2N} g^2 - 2U)^{N-1}} =: 2J_\perp,
\end{align}
where we have used the relation the relation
\begin{align}\label{eq:VN-matrix-element}
  \<N0|&V^N|0N\> \span\nonumber\\
  &=\<N0|c^{\alpha_1\dagger} U^{\alpha_1\beta_1} d^{\beta_1} \cdots c^{\alpha_N\dagger} U^{\alpha_N\beta_N} d^{\beta_N} |0N\> \nonumber\\
  &= \varepsilon^{\alpha_1\cdots\alpha_N} \varepsilon^{\beta_1\cdots\beta_N}\<N0| U^{\alpha_1\beta_1} \cdots U^{\alpha_N\beta_N} |N0\> \nonumber\\
  &= \frac{1}{N!} \varepsilon^{\alpha_1\cdots\alpha_N} \varepsilon^{\beta_1\cdots\beta_N} \varepsilon_{\alpha_1\cdots\alpha_N} \varepsilon_{\beta_1\cdots\beta_N} = N!.
\end{align}
Therefore, to the leading order in $t$, the effective Hamiltonian of our qubit model in the 4d space can be written as an $XXZ$ spin chain,
\begin{align}
  H_\eff = J_\perp (X_1 X_2 + Y_1 Y_2) + J_z (Z_1 Z_2 - 1).
\end{align}

We remark that this calculation is also valid in the traditional lattice model. This is because in this calculation there is no use of the truncation in the gauge link Hilbert space. Even in the derivation of \cref{eq:VN-matrix-element}, the gauge link is automatically restricted to the anti-symmetric subspace in each intermediate state by the anti-commuting nature of the fermions.

\subsection{The dimer phase}\label{app:dimer}
In this appendix, we calculate the leading order correction to the entanglement entropy in the dimer phase due to the hopping term $t$. As we argued in \cref{subsec:dimer}, when $U/t \gg 1$ and $g^2>0$, the ground state is dimerized, and when $N$ is even, the ground state is doubly degenerate. For simplicity, we will only consider the case of $N$ even in the following. When $N$ is odd, it can be analyzed in a similar way.

In $U/t \rightarrow \infty$ limit, the ground state was shown to be a tensor product of dimers in \cref{eq:SUN-dimer}. When $U/t$ is large but finite, the ground state becomes a superposition of the dimer product state with other states. However, the EE is still extremely localized, and it is sufficient to only focus on a single dimer. As we saw in the main text, even with this simplification, our calculation still leads to quantitative correction to the EE in perturbation of $t/U$.

For conciseness, We will label our dimer state
\begin{align}
  \frac{1}{\binom{N}{N/2}} \Big(|\alpha_1 & \cdots \alpha_{\frac{N}{2}}\>_{2i-1} |D^{[\frac{N}{2}]}_{[\alpha_1 \cdots \alpha_{\frac{N}{2}}][\gamma_1 \cdots \gamma_{\frac{N}{2}}]}\>_{2i-1, 2i} \nonumber\\
  &\varepsilon^{\gamma_1 \cdots \gamma_{\frac{N}{2}}\beta_1 \cdots \beta_{\frac{N}{2}}}|\beta_1 \cdots \beta_{\frac{N}{2}}\>_{2i}\Big) \tag{re \ref{eq:SUN-dimer}}
\end{align}
using the fermion number of the two sites, denoted as $|\frac{N}{2}\frac{N}{2}\>$. The gauge link can be determined by Gauss's law, and thus is omitted in the state. The hopping term will mix this state with $|\frac{N}{2}+1,\frac{N}{2}-1\>$ and $|\frac{N}{2}-1,\frac{N}{2}+1\>$,
\begin{align}
  \<\frac{N}{2}+1,\frac{N}{2}-1| V |\frac{N}{2}\frac{N}{2}\> = tu,
\end{align}
where $u$ is a real factor that depends on the matrix element of $U^{\alpha\beta}$ between the link state $|D^{[\frac{N}{2}]}\>$ and $|D^{[\frac{N}{2}+1]}\>$, and $u = \sqrt{2}$ for $N=2$. Similarly, we have 
\begin{align}
  \<\frac{N}{2}-1,\frac{N}{2}+1| V |\frac{N}{2}\frac{N}{2}\> = tu,
\end{align}
where $u$ remains the same due to charge conjugation symmetry. Furthermore, we have
\begin{align}
  H_0 |\frac{N}{2}\frac{N}{2}\> &= \frac{1}{8} g^2 N(N+1) + \frac{1}{2} U N^2 \\
  H_0 |\frac{N}{2} \pm 1,\frac{N}{2} \mp 1\> &= \frac{N+1}{8N} g^2 (N^2-4) + \frac{1}{2} U (N^2-4). \nonumber
\end{align}
Then to the first order in the perturbation $t/U$, the ground state $|0\>$ can be written as
\begin{align}
  |0\> = |\frac{N}{2}\frac{N}{2}\> + \frac{tu}{\frac{N+1}{2N}g^2 + 2U} \Big( |\frac{N}{2}+1,\frac{N}{2}-1\> \nonumber\\
  + |\frac{N}{2}-1,\frac{N}{2}+1\>\Big).
\end{align}
To the leading order in $t/U$, we can set $g^2=0$.

The reduced density matrix of the first fermion site has eigenvalues only in the three sectors $|\frac{N}{2}\>$ and $|\frac{N}{2} \pm 1\>$, which are $\big(\<0|0\> \binom{N}{N/2}\big)^{-1}$ and $(\frac{tu}{2U})^2 \big(\<0|0\> \binom{N}{N/2-1}\big)^{-1}$. Therefore, the von Neumann EE can be calculated as
\begin{align}
  &\frac{1}{\<0|0\>} \Bigg( \ln \left(\<0|0\> \binom{N}{N/2}\right) \nonumber\\[1\jot]&\qquad\quad+ 2\big(\frac{tu}{2U}\big)^2 \ln \big(\frac{tu}{2U}\big)^{-2} \left(\<0|0\> \binom{N}{N/2-1}\right) \Bigg) \nonumber\\
  &= \ln\binom{N}{N/2} + (\frac{tu}{U})^2 \ln \frac{2U}{tu} + O\big((\frac{t}{U})^2\big).
\end{align}
Similarly, we can calculate the EE between the dimers to be
\begin{align}
  \big(\frac{tu'}{U}\big)^2 \ln \frac{2U}{tu'} + O\big((\frac{t}{U})^2\big),
\end{align}
where $u' = \frac{1}{\sqrt{2}}$ for $N=2$. (Notice that in this case there are no extra factors for normalizing the states, because the contractions of indices are simply shifted from fermions to links.) Therefore for $N=2$, we find that the difference of the EE is
\begin{align}
  \Delta S = \ln 2 + \frac{3}{2}\big(\frac{t}{U}\big)^2 \ln U + O\Big(\big(\frac{t}{U}\big)^2\Big).
\end{align}

\section{Generation of the electric field term}\label{app:generation}
According to Wilson's \ac{RG}, any term that respects the symmetry of the model may be generated under \ac{RG}. From the perspective of a lattice model, this means that even when the electric field term with coupling $g^2$ is absent, it might be generated by other terms that respect the same symmetry. In this appendix, we will show that the qubit model is different from the traditional model from this perspective: even though $g^2=0$ in the qubit model, it can effectively be generated by the fermion hopping term, while this is not the case in the traditional model.

Since it is difficult to perform Wilson's \ac{RG} analysis for the full model, our arguments will be limited to the single link model with the Hamiltonian 
\begin{align}\label{eq:H-one-hop}
  H = c^{\alpha\dagger} U^{\alpha\beta} d^\beta + \hc,
\end{align}
where we have suppressed the site indices and used $c$ and $d$ to distinguish the fermion operators on the two sites connected by a link. We can then compute the effective Hamiltonian for the gauge field by taking a partial trace of the partition function over the fermion Hilbert space
\begin{align}
\label{eq:H-one-hop-eff}
  H_\gauge := -\frac{1}{\beta} \ln(\tr_f\e^{-\beta H}).
\end{align}
Note that the trace here is performed 
on the full Hilbert space and not restricted to the gauge invariant subspace, since we would like to understand how the effective gauge Hamiltonian depends on the different Gauss law sectors. This is important because the effective gauge Hamiltonian can in principle combine with neighboring links to form new types of effective gauge invariant Hamiltonians. Below we will argue that in the traditional theory, $H_\gauge^\mathrm{trad}$ is proportional to the identity operator, while in our qubit model, $H_\gauge^\mathrm{qubit}$ is proportional to the electric field operator.

\subsection{Traditional model}

In the traditional model, \cref{eq:H-one-hop} is defined on the infinite-dimensional link Hilbert space as explained in \cref{sec:qubit}. So effective Hamiltonian in the traditional model $H^\mathrm{trad}_\gauge$ will be an operator in this infinite dimensional Hilbert space. Let us find its matrix elements in the ``position basis'' $|g\>: g \in G$. From 
\cref{eq:uong} we know that they are eigenstates of the link operators $U^{\alpha\beta}$ operator with eigenvalues $D^f_{\alpha\beta}(g)$. Therefore 
\begin{align}
  (\tr_f\e^{-\beta H}) |g\> = \tr_f\e^{-\beta H(g)} |g\>,
\end{align}
where
\begin{align}
  H(g) = c^{\alpha\dagger} D^f_{\alpha\beta}(g) d^\beta + \hc.
\end{align}
Thus, the basis vecgtors $|g\rangle$ are also eigenstates of $H^\mathrm{trad}_\gauge$.

Let us now show that $\tr_f\e^{-\beta H(g)}$ is in fact independent of $g$. Consider 
\begin{align}
  H(gh) &= c^{\alpha\dagger} D^f_{\alpha\beta}(gh) d^\beta + \hc \nonumber\\
  &= c^{\alpha\dagger} D^f_{\alpha\gamma}(g) D^f_{\gamma\beta}(h) d^\beta  + \hc. 
\end{align}
$D^f_{\gamma\beta}(h) d^\beta$ is a unitary transformation of the fermionic $d$ operators, and from the Stone–von Neumann theorem for fermions, we know that there exists a unitary operator $\mathcal{U}(h)$ on the fermion Hilbert space that implements this transformation through the relation
\begin{align}
  D^f_{\gamma\beta}(h) d^\beta = \mathcal{U}(h)\  d^\gamma \ \mathcal{U}(h)^{-1}.
\end{align}
This $\mathcal{U}(h)$ operator will cancel when tracing out the fermions, and we get
\begin{align}
  \tr_f\e^{-\beta H(gh)} = \tr_f\e^{-\beta H(g)}
\end{align}
for arbitrary $h \in G$. This means the eigenvalues $\tr_f\e^{-\beta H(g)}$ are independent of $g$ implying that
\begin{align}
  H_\gauge^\mathrm{trad} \propto \mathbbm{1}_\gauge ,
\end{align}
where $\mathbbm{1}_\gauge$ is the identity operator on the link Hilbert space.

The above result is not surprising since we already showed in \cref{subsec:KS-Hamiltonian}, that when $g^2=0$ we can perform a unitary transformation on the fermion operators and convert the traditional Hamiltonian into a free-fermion Hamiltonian. Since the free-fermion Hamiltonian has a larger symmetry ($\O(2N)$) than the gauge theory, which only has the symmetry $\U(1) \times \SU(N)$ or $\SU(2)_c \times \SU(2)_s$ when $N=2$, the electric field term cannot be generated.

\subsection{Qubit model}

Unlike the traditional model, in the qubit model the fermion hopping term does not have higher symmetry than the electric field term, and we will show below that the electric field term is indeed generated by the hopping term. In this case, the partial trace can be expanded as
\begin{align}
  \tr_f\e^{-\beta H} = \sum_{n=0}^\infty \frac{1}{n!} (-\beta)^n \tr_f(H^n).
\end{align}
In order to have non-vanishing $\tr_f(H^n)$, $H^n$ must preserve the fermion numbers of each color on each site, i.e., the number of $c^\alpha$ should equal to the number of $c^{\alpha\dagger}$ in $H^n$ and similarly for $d$, which in turn implies that the number of $U$ should equal to $U^\dagger$. As a result, $\tr_f(H^n)$ is non-vanishing only for $n$ even.

Since we no longer have a basis of simultaneous eigenstates for all the $U^{\alpha\beta}$ operators, we have to calculate the fermionic trace of each term in the expansion by inserting the link operators explicitly. In the following, we will calculate $H_\gauge^\mathrm{qubit}$ for $N=2$ and $3$ as examples. 

\subsubsection{\texorpdfstring{$\SU(2)$}{SU(2)}}

To simplify the calculations we introduce the shorthand notation for the basis states of the link Hilbert space as $|0\rangle$ and $|\alpha\beta\> := |D^f_{\alpha\beta}\>$. In this notation can explicitly write the link matrices as
\begin{align}
  \Ul^{\alpha\beta} = \frac{1}{\sqrt{2}}(|\alpha\beta\>\<0| + \varepsilon_{\alpha\alpha'}\varepsilon_{\beta\beta'}|0\>\<\alpha'\beta'|).
\end{align}
Let us first consider 
\begin{align}
  \tr_f(H^2) = \tr\big(c^{\alpha\dagger} U^{\alpha\beta} d^\beta + d^{\beta\dagger} (U^{\alpha\beta})^\dagger c^\alpha\big)^2.
\end{align}
From our earlier discussion, we know that we only need to consider the cross terms, and the number of $c^{\alpha\dagger}$ and $c^\alpha$ (resp. $d^\dagger$ and $d$) should be equal. There are two such terms, one of which takes the form
\begin{align}
  c^{\alpha\dagger} U^{\alpha\beta} d^\beta d^{\beta\dagger} (U^{\alpha\beta})^\dagger c^\alpha = 
  \frac{n^\alpha (1-n^\beta)}{2}  (|\alpha\beta\>\<\alpha\beta| + |0\>\<0|),
\end{align}
where the repeated indices are not summed over. Since the fermion sites are uncorrelated, when tracing out the fermions, we can simply replace $n_i^\alpha$ by its expected value $\<n_i^\alpha\> = \frac{1}{2}$. Therefore we have
\begin{align}
  \frac{1}{16}\tr_f(c^{\alpha\dagger} U^{\alpha\beta} d^\beta d^{\beta\dagger} (U^{\alpha\beta})^\dagger c^\alpha) = \frac{1}{8} (|\alpha\beta\>\<\alpha\beta| + |0\>\<0|) ,
\end{align}
Summing over $\alpha$ and $\beta$ and adding the other cross term, we have
\begin{align}
  \tr_f(H^2) = 16|0\>\<0| + 4 \sum_{\alpha,\beta}|\alpha\beta\>\<\alpha\beta| .
\end{align}
Using a Mathematica, we can show that
\begin{align}
  \tr_f(H^{2n}) &= (4^n+12) \Big(|0\>\<0| + \frac{1}{4} \sum_{\alpha,\beta}|\alpha\beta\>\<\alpha\beta|\Big) ,
\end{align}
for $n\geq 1$. Therefore 
\begin{align}
  &\quad \tr_f\e^{-\beta H} = \sum_{n=0}^\infty \frac{1}{(2n)!} \beta^{2n} \tr_f(H^{2n}) \nonumber\\
  &= \sum_{n=0}^\infty \frac{1}{(2n)!} \beta^{2n} (4^n+12) \Big(|0\>\<0| + \frac{1}{4} \sum_{\alpha,\beta}|\alpha\beta\>\<\alpha\beta|\Big) \nonumber\\
  &\quad + 3|0\>\<0| + \frac{51}{4} \sum_{\alpha,\beta}|\alpha\beta\>\<\alpha\beta| \nonumber\\
  &= (\cosh2\beta + 12\cosh\beta) \Big(|0\>\<0| + \frac{1}{4} \sum_{\alpha,\beta}|\alpha\beta\>\<\alpha\beta|\Big) \nonumber\\
  &\quad + 3|0\>\<0| + \frac{51}{4} \sum_{\alpha,\beta}|\alpha\beta\>\<\alpha\beta|.
\end{align}
For large $\beta$ we have
\begin{align}
  \beta H_\gauge^\mathrm{qubit} &\approx 2\ln2 \sum_{\alpha,\beta}|\alpha\beta\>\<\alpha\beta| + (\ln2 - 2\beta) \mathbb{I} \nonumber\\
  &=\frac{4}{3} \ln2 (L^{a2} + R^{a2}) + (\ln2 - 2\beta) \mathbb{I},
\end{align}
which shows that the electric field term $(L^{a2} + R^{a2})$ is indeed generated.

\subsubsection{\texorpdfstring{$\SU(3)$}{SU(3)}}

We can repeat the above calculations for $N=3$. We now introduce the shorthand notation for the basis states of the link Hilbert space as $|0\rangle$ and $|\alpha\beta\> := |D^f_{\alpha\beta}\>$ and $|\overline{\alpha\beta} \> := |D^{\bar{f}}_{\alpha\beta}\>$. In this case we have
\begin{align}
  \Ul^{\alpha\beta} &= \frac{1}{\sqrt{3}}(|\alpha\beta\>\<0| + |0\>\<\overline{\alpha\beta}|) \nonumber \\
  &\qquad + \frac{1}{2} \varepsilon_{\alpha\gamma\gamma'}\varepsilon_{\beta\delta\delta'}|\overline{\gamma'\delta'}\>\<\gamma\delta|).
\end{align}
Similar to the $\SU(2)$ case, for fixed $\alpha$ and $\beta$, the two crossing terms are
\begin{align}
  &\quad c^{\alpha\dagger} U^{\alpha\beta} d^\beta d^{\beta\dagger} (U^{\alpha\beta})^\dagger c^\alpha \nonumber\\
  &= n^\alpha (1-n^\beta) \Big(\frac{1}{3} (|\alpha\beta\>\<\alpha\beta| + |0\>\<0|) + \frac{1}{4} \sum_{\substack{\gamma\neq\alpha\\\delta\neq\beta}}|\overline{\gamma\delta}\>\<\overline{\gamma\delta}| \Big), \nonumber\\
  &\quad d^{\beta\dagger} (U^{\alpha\beta})^\dagger c^\alpha c^{\alpha\dagger} U^{\alpha\beta} d^\beta \\
  &= (1-n^\alpha) n^\beta \Big(\frac{1}{3} (|0\>\<0| + |\overline{\alpha\beta}\>\<\overline{\alpha\beta}| ) + \frac{1}{4} \sum_{\substack{\gamma\neq\alpha\\\delta\neq\beta}}|\gamma\delta\>\<\gamma\delta| \Big). \nonumber
\end{align}
Again, replacing $n_i^\alpha$ by its expectation value, summing over $\alpha$ and $\beta$ and adding the two cross terms, we get
\begin{align}
  \tr_f(H^2) &= 96|0\>\<0| + \frac{64}{3} \sum_{\alpha,\beta}(|\alpha\beta\>\<\alpha\beta|+|\overline{\alpha\beta}\>\<\overline{\alpha\beta}|) .
\end{align}
Again using Mathematica, we can show
\begin{align}
  \tr_f(H^{2n}) &= a_n|0\>\<0| \nonumber \\
  & \qquad + b_n \sum_{\alpha,\beta}(|\alpha\beta\>\<\alpha\beta|+|\overline{\alpha\beta}\>\<\overline{\alpha\beta}|) ,
\end{align}
where
\begin{align}
  \lim_{n\rightarrow\infty} \frac{a_n}{b_n} = 6.
\end{align}
Therefore, when $\beta$ is large, we obtain
\begin{align}
  \beta H_\gauge^\mathrm{qubit} &\approx \ln6 \sum_{\alpha,\beta}(|\alpha\beta\>\<\alpha\beta|+|\overline{\alpha\beta}\>\<\overline{\alpha\beta}|) + c \mathbb{I} \nonumber\\
  &=\frac{3}{8} \ln6 (L^{a2} + R^{a2}) + c \mathbb{I},
\end{align}
where $c$ is a constant. Again, we see that the electric field term $(L^{a2} + R^{a2})$ is generated.

\section{Infrared and ultraviolet central charge}\label{app:central-charge}
In a 2d CFT (see Ref.~\cite{Ginsparg:1988ui} for a review), the generators of the conformal symmetry can be written in terms of the stress-energy tensor. Working in light cone coordinates, \(z=x+it\) in Euclidean metric, the stress-energy tensor is completely determined by its holomorphic (left mover) and antiholomorphic (right mover) parts, which are denoted by\footnote{By conservation of the stress-energy tensor and its tracelessness, \(T_{z\bar z}=T_{\bar z z}=0\), that follows from conformal invariance, \(T_{zz}\) has no dependence on \(\bar z\) nor \(T_{\bar z\bar z}\) on \(z\).} \(T_{zz}\equiv T(z)\) and \(T_{\bar z\bar z} \equiv\overline T(\bar z)\), respectively.  On general grounds, we have the operator product expansion\footnote{Exactly analogous expressions can be written for the antiholomorphic part of the stress-energy tensor, but we skip the details of the discussion since conservation of stress-energy tensor in a Lorentz symmetric theory restricts the holomorphic and the antiholomorphic central charges to be the same.}
\begin{equation}
    T(z) T(w) = \frac {c/2}{(z-w)^4} + \frac {2T(w)}{(z-w)^2} + \frac{\partial T(w)}{z-w} + \cdots \,.\label{eq:Tcorr}
\end{equation}
If we expand \(T(z) = \sum_n z^{-n-2} L_n\), then the operators \(L_n\) generate a Virasoro algebra with a central extension:
\begin{equation}
[L_m,L_n] = (m-n) L_{n+m} + \delta_{m,-n} \frac{(m-1)m(m+1)}{12} c\,,
\end{equation}
which justifies the name central charge for the constant \(c\).  From \cref{eq:Tcorr}, it is clear that each free degree of freedom contributes a constant amount to the central charge, the normalization chosen is such that this is \(c=1\) for each free real bosonic degree of freedom and each free Dirac fermionic degree. Even though in the rest of this discussion we treat the central charge as counting the number of degrees of freedom, if the fields are interacting this should be understood as the equivalent number of free-fields and is, thus, not constrained to be an integer.  All that is known in general is that non-negativity of the Hilbert space norm implies either \(c\geq1\) or \(c=1 - \frac6{m(m+1)}\) for some\footnote{The case \(m=2\), i.e., \(c=0\), gives a trivial Hilbert space.\\[\baselineskip]}  integer \(m\geq2\).

The entanglement between a region and its outside in a CFT also counts the number of degrees of freedom. After regulating the theory in both the infrared and the ultraviolet, each mode of each field contributes to the entanglement.  This picture predicts that the entanglement entropy should, therefore, grow as the logarithm of the number of modes that can be accommodated between the ultraviolet and the infrared cutoffs, with a coefficient proportional to the number of contributing fields.  In fact, an explicit calculation~\cite{Holzhey:1994we,Calabrese:2004eu,Calabrese:2009qy} shows that in a CFT, the coefficient of the logarithm is \(\frac c3\) if the region is an interior region (and thus separated from the outside by both a left and a right boundary), as we presented in \cref{eq:cIR}.  If, on the other hand, the theory has a mass gap, then the entanglement falls off exponentially with distance.  As a result, as the system size is increased far beyond the largest correlation length, the entanglement entropy saturates and no longer displays the logarithmic dependence on the system size.  Thus, the central charge determined using \cref{eq:cIR} essentially counts the degrees of freedom that are massless (or have masses much less than the box size) and is, thus, a measure of the infrared properties of the theory.  In our work, we denote this by using the symbol \(c_\IR\).  In a theory with a mass gap, the lack of massless modes and the resulting saturation of the entanglement entropy is expressed by saying \(c_\IR=0\) when measured on sufficiently large boxes. In practical calculations, when the box size is not large enough, we can sometimes get $c_\IR \neq 0$. It is also important to realize that there can be cross-over phenomena from one fixed point to another in asymptotically free theories as recently discovered in qubit regularized models \cite{Bhattacharya:2020gpm,Maiti:2023kpn}. In such cases $c_\IR$ can vary from one value to another while ultimately vanishing.

On the other hand, the value at which the entanglement entropy saturates for large system sizes grows as the logarithm of the number of modes between the largest correlation length and the ultraviolet cutoff.  Specifically, we can consider the CFT that arises when the correlation length grows to infinity as we approach a critical point of the lattice theory, whether or not the continuum field theory that describes the rescaled physics of the low modes is conformal.  If we consider the entanglement entropy between the left and right halves of an infinitely large system (i.e., between two regions separated by a single boundary), a direct calculation in the CFT shows that the coefficient of the logarithm of the correlation length is given by \(\frac c6\) as presented in \cref{eq:cUV}.  When measured sufficiently close to the critical point, this measure of the central charge essentially counts the entirety of the lattice degrees of freedom that contribute to the continuum field theory by becoming massless compared to the lattice scale, and, thus, is an ultraviolet property of the continuum field theory. We make this explicit by denoting this central charge by \(c_\UV\). Again, away from the critical point, $c_\UV$ calculated using \cref{eq:cUV} is only an effective measure of the number of fields whose correlation lengths are growing proportionally as we approach the critical point. If the correlation lengths grow at different rates, which can surely occur, we can get values effective values of $c_\UV$ that do not correspond to any fixed point.

\begin{figure}[htb]
  \centering
  \includegraphics[width=0.8\linewidth]{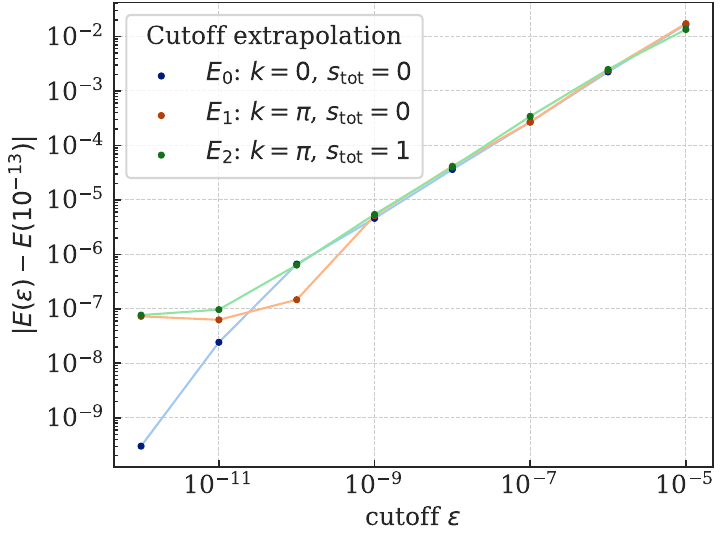}
  \caption{Errors in energies of the lowest $3$ states as functions of cutoff $\varepsilon$ at $g^2=0$, $U=0.1$ and $L=16$ with periodic boundary conditions.}
  \label{fig:cutoff-extrapolation}
\end{figure}

\section{Numerical methods}
\label{app:ITensor}

In the numerical study, we use the DMRG algorithm provided by the ITensor \cite{ITensor,ITensor-r0.3} package to obtain the low-lying states. We do not set an upper bound to the bond dimension. Instead, the bond dimension is controlled by the truncation of the singular values. In general, a finite cutoff of the singular values will introduce errors, which can be estimated by the energy eigenvalues. In \cref{fig:cutoff-extrapolation}, we plot the estimated errors of the lowest $3$ states as a function of the cutoff at $g^2=0$, $U=0.1$ and $L=16$ with periodic boundary conditions. The errors are determined by the difference of energies at cutoff $\varepsilon$ and at cutoff $10^{-13}$. From the figure, we see that in the logarithmic scale, the errors roughly scale linearly with the cutoff, and they are quite small when $\varepsilon = 10^{-10}$. In order to be more accurate, we chose $\varepsilon = 10^{-12}$ in all our simulations.

\section{Data table for the critical points}
\label{app:data}

In \cref{tab:Uc}, we present the critical values \(U_c\) for various values of \(g^2\) and \(L\), and the extrapolation to $L \rightarrow \infty$. In \cref{tab:S-xi}, we present the correlation length $\xi$ and the entanglement entropy $S$ at various $g^2$ and $U$.

\begin{table*}[htb]
  \centering
  \caption{The critical points $U_c(g^2, L)$ and extrapolated value $U_c(g^2, \infty)$}
  \begin{tabular}{S[table-format=2.0] *{16}{S[table-format=2.7]}}
    \toprule
    \multicolumn{1}{c}{\multirow{2}{*}{$L$}} & \multicolumn{8}{c}{$g^2$} \\
    \cmidrule(lr){2-9}
    \multicolumn{1}{c}{} & 
    \multicolumn{1}{c}{$-1$} &        \multicolumn{1}{c}{$-0.7$} &        \multicolumn{1}{c}{$-0.5$} &        \multicolumn{1}{c}{$-0.4$} &        \multicolumn{1}{c}{$-0.3$} &        \multicolumn{1}{c}{$-0.2$} &        \multicolumn{1}{c}{$-0.1$} &           \multicolumn{1}{c}{$0$} \\
\midrule
 4 & 0.19389(3) & 0.11028(2) & 0.07468(2) & 0.06408(3) & 0.05829(2) & 0.05707(3) & 0.06002(2) & 0.06664(3) \\
 6 & 0.17968(6) & 0.10995(4) & 0.08247(4) & 0.07472(4) & 0.07084(4) & 0.07059(4) & 0.07369(3) & 0.07975(4) \\
 8 & 0.17389(5) & 0.10942(4) & 0.08448(5) & 0.07771(4) & 0.07441(5) & 0.07451(5) & 0.07774(4) & 0.08367(6) \\
10 & 0.17103(7) & 0.10883(6) & 0.08516(6) & 0.07875(6) & 0.07582(6) & 0.07618(5) & 0.07935(6) & 0.08534(7) \\
12 & 0.16934(8) & 0.10845(6) & 0.08545(6) & 0.07928(6) & 0.07651(6) & 0.07698(6) & 0.08035(5) & 0.08619(6) \\
$\infty$ & 0.1643(2) & 0.10800(6) & 0.08594(5) & 0.08014(8) & 0.07765(1) & 0.07836(5) & 0.0818(1) & 0.08770(4) \\
\toprule
\multicolumn{1}{c}{\multirow{2}{*}{$L$}} & \multicolumn{8}{c}{$g^2$} \\
\cmidrule(lr){2-9}
\multicolumn{1}{c}{} &         
\multicolumn{1}{c}{$0.1$} &         \multicolumn{1}{c}{$0.2$} &         \multicolumn{1}{c}{$0.5$} &           \multicolumn{1}{c}{$1$} &           \multicolumn{1}{c}{$2$} &           \multicolumn{1}{c}{$3$} &           \multicolumn{1}{c}{$4$} &           \multicolumn{1}{c}{$5$} \\
\midrule
 4 & 0.07652(3) & 0.08923(6) & 0.14035(9) &   0.25500(4) & 0.53877(8) & 0.85654(5) & 1.19123(7) & 1.53567(5) \\
 6 & 0.08858(5) &  0.09970(4) & 0.14502(4) & 0.24848(4) & 0.51326(5) &  0.81720(5) & 1.14169(6) & 1.47859(7) \\
 8 & 0.09222(6) & 0.10299(5) & 0.14661(5) & 0.24665(5) & 0.50509(6) & 0.80418(7) &  1.12510(9) & 1.45901(7) \\
10 &  0.09390(8) & 0.10452(8) & 0.14741(8) & 0.24587(6) & 0.50145(8) & 0.79834(9) & 1.11746(9) & 1.45016(9) \\
12 & 0.09458(7) & 0.10513(7) &  0.14770(6) & 0.24545(7) & 0.49952(9) &  0.7952(1) & 1.11328(8) &  1.4454(1) \\
$\infty$ & 0.0961(1) & 0.10658(9) & 0.14857(9) & 0.24485(6) & 0.4958(1) & 0.78887(6) & 1.1047(2) & 1.43500(9) \\
\bottomrule
  \end{tabular}
  \label{tab:Uc}
\end{table*}

\begin{table*}[htb]
  \centering
  \caption{Correlation length $\xi$ and entanglement entropy $S$ at various $g^2$ and $U$}
  \begin{tabular}{cp{1em} *{13}{l}}
    \toprule
    \multirow{3}{*}[-0.4em]{$g^2 = -0.5$} & $U$ & 0.8 & 0.9 & 1 & 1.1 & 1.2 & 1.3 & 1.4 & 1.5 & 1.6 & 1.7 & 1.8 & 1.9 & 2 \\
    \cmidrule(r){2-15}
    & $\xi$ & 26.528 & 16.486 & 10.98 & 7.812 & 5.883 & 4.63 & 3.779 & 3.169 & 2.704 & 2.359 & 2.088 & 1.864 & 1.69 \\
    \cmidrule(r){2-15}
    & $S$ & 1.1471 & 1.0133 & 0.899 & 0.8028 & 0.7211 & 0.651 & 0.5903 & 0.5375 & 0.4911 & 0.4503 & 0.4141 & 0.382 & 0.3533 \\
    \toprule
    \multirow{3}{*}[-0.4em]{$g^2 = 0$} & $U$ & 0.7 & 0.8 & 0.9 & 1 & 1.1 & 1.2 & 1.3 & 1.4 & 1.5 & 1.6 & 1.7 & 1.8 \\
    \cmidrule(r){2-15}
    & $\xi$ & 49.14 & 29.34 & 18.32 & 12.11 & 8.54 & 6.36 & 4.97 & 4.02 & 3.36 & 2.88 & 2.53 & 2.26 \\
    \cmidrule(r){2-15}
    & $S$ & 1.2594 & 1.1139 & 0.9787 & 0.8626 & 0.7655 & 0.6839 & 0.6147 & 0.5553 & 0.5040 & 0.4594 & 0.4204 & 0.3861 \\
    \toprule
    \multirow{3}{*}[-0.4em]{$g^2 = 2$} & $U$ & 1 & 1.1 & 1.2 & 1.3 & 1.4 & 1.5 & 1.6 & 1.7 & 1.8 & 1.9 \\
    \cmidrule(r){2-15}
    & $\xi$ & 72.339 & 45.792 & 27.292 & 16.949 & 11.256 & 7.99 & 6.006 & 4.724 & 3.846 & 3.22 \\
    \cmidrule(r){2-15}
    & $S$ & 1.2164 & 1.0849 & 0.9356 & 0.7974 & 0.6826 & 0.5895 & 0.5137 & 0.4515 & 0.3999 & 0.3569 \\
    \toprule
    \multirow{3}{*}[-0.4em]{$g^2 = 5$} & $U$ & 2 & 2.1 & 2.2 & 2.3 & 2.4 & 2.5 & 2.6 \\
    \cmidrule(r){2-15}
    & $\xi$ & 51.303 & 29.489 & 17.737 & 11.503 & 8.027 & 5.968 & 4.659 \\
    \cmidrule(r){2-15}
    & $S$ & 1.0309 & 0.8731 & 0.7233 & 0.5992 & 0.5002 & 0.4214 & 0.3582 \\
    \bottomrule
  \end{tabular}
  \label{tab:S-xi}
\end{table*}

\end{document}